\pdfoutput=1
\documentclass[a4paper]{amsart}
\usepackage[utf8]{inputenc}
\usepackage[english]{babel}
\usepackage[margin=2cm]{geometry}

\usepackage{graphicx}
\usepackage{siunitx}

\let\vec\mathbf %Bold vectors

\title{Creation of equal-spin triplet superconductivity at the Al/EuS interface}

\author{S. Diesch$^{1}$, P. Machon$^{1}$, M. Wolz$^{1}$, C. Sürgers$^{2}$,
  D. Beckmann$^{3}$, W. Belzig$^{1*}$ and E. Scheer$^{1*}$}

\begin{document}

\begin{abstract}
In conventional superconductors, electrons of opposite spins are bound into Cooper pairs. However, when the superconductor is in contact with a non-uniformly ordered ferromagnet, an exotic type of superconductivity can appear at the interface, with electrons bound into three possible spin-triplet states. Triplet pairs with equal spin play a vital role in low-dissipation spintronics. Despite the observation of supercurrents through ferromagnets, spectroscopic evidence for the existence of equal-spin triplet pairs is still missing. Here we show a theoretical model that reveals a characteristic gap structure in the quasiparticle density of states which provides a unique signature for the presence of equal-spin triplet pairs. By scanning tunnelling spectroscopy we measure the local density of states to reveal the spin configuration of triplet pairs. We demonstrate that the Al/EuS interface causes strong and tunable spin-mixing by virtue of its spin-dependent transmission.
\end{abstract}

\maketitle

\stepcounter{footnote}\footnotetext{Department of Physics, University of Konstanz,
  Universitätsstraße 10, D-78457 Konstanz, Germany}
\stepcounter{footnote}\footnotetext{Physikalisches Institut, Karlsruhe Institute of
  Technology (KIT), Wolfgang Gaede Straße 1, D-76131 Karlsruhe, Germany}
\stepcounter{footnote}\footnotetext{Institute of Nanotechnology, Karlsruhe Institute of
  Technology (KIT), Hermann-von-Helmholtz-Platz 1, D-76344 Eggenstein-Leopoldshafen,
  Germany}

Hybrid superconductor-ferromagnet (S/F) heterostructures are fundamental building blocks for next-generation, ultralow power computers. In these devices, Cooper pairs of equal spin would carry spin information without dissipation \cite{linder2015, eschrig2011}, reducing power consumption by several orders of magnitude \cite{holmes2013}.
Due to these promising applications, such exotic electron states have attracted considerable interest in recent years: After the initial prediction of triplet superconductivity in conventional s-wave superconductors, the theory has been elaborated to cover several scenarios including multilayers and microscale devices \cite{balatsky1992, bergeret2001, bergeret2001b, kadigrobov2001, eschrig2003,  bergeret2005, eschrig2015}. The Pauli principle can be fulfilled for both singlet and
triplet Cooper pairs by adjusting the symmetry in the time argument.
Thus, spin triplet pairs in s-wave superconductors with even spatial symmetry must have a pair correlation function that is
odd in the time argument, called odd frequency spin triplets
\cite{bergeret2001, eschrig2008}. Triplet pairs are created at interfaces between superconductors and
ferromagnetic materials, and their properties have been studied theoretically intensively \cite{buzdin2005, asano2006}.
When a single magnetisation direction is present, solely $m = 0$
triplet pairs ($\frac{1}{\sqrt{2}} [|\uparrow \downarrow \,\rangle + |\downarrow
\uparrow \, \rangle]$) are created by spin-mixing, i.e., spin-dependent phase shifts for the electrons scattering at this interface \cite{eschrig2008}. These triplet pairs can be converted into Cooper pairs with parallel spins ($m = \pm1, | \uparrow\uparrow \rangle / | \downarrow\downarrow
\rangle$) by a second, noncollinear magnetisation direction defining a new spin quantisation axis \cite{eschrig2015}.
 We will refer to
these different kinds of pairings as mixed-spin ($m = 0$) or equal-spin ($m = \pm 1$)
state.

Experimentally, the existence of equal-spin pairing has been shown
by measuring the penetration depth of the supercurrent in ferromagnetic
materials \cite{keizer2006, khaire2010, robinson2010}, where the term long-ranged
spin-triplet superconductivity has been coined.  More experimental indication for spin-triplet superconductivity has been reported from measurements of the
critical temperature on S/F/F' spin valves \cite{leksin2012, jara2014, flokstra2015, singh2015}, from measurements of the supercurrent through S/F/F'/F''/S  junctions as a function of the relative magnetisation directions of the three ferromagnetic layers \cite{gingrich2010, martinez2016} and by studies of the paramagnetic Meissner effect in S/F bilayers \cite{dibernardo2015b}. These works, however, do not consider noncollinear magnetisations and can therefore, in principle, not distinguish different types of spin-triplet superconductivity. Obtaining experimental support for the creation of triplet pairs from measurements of the quasiparticle density of states by scanning tunnelling spectroscopy (STS) has been suggested \cite{linder2010,ouassou2017}. In a previous study, STS has been performed on Nb superconducting films proximity coupled to the chiral ferromagnet Ho \cite{bernardo2015}. In that work a Nb/Ho bilayer has been probed from the superconducting side and the obtained spectra were compared to a specialised theoretical model relying on the chiral magnetic state of Ho and pinning effects occurring at the Nb/Ho interface. Both zero-bias peaks and double peaks in the spectra were observed and interpreted as signatures of triplet superconductivity consistent with theory \cite{ouassou2017}, however, only very qualitative agreement between experimental and theoretical spectra could be achieved, most likely because the full boundary conditions were not used \cite{eschrig2015-2}.

As a result, it has not been revealed whether these triplets are equal-spin or mixed-spin pairs due to a missing unique signature which would allow to distinguish between these two possibilities experimentally. Furthermore, the influence of the spin-dependent phase shifts are expected to be much weaker on the S side of the interface, resulting in only small amplitudes of the subgap features.
To address these two issues, we study here the local density of states (LDOS) of an S/FI/N trilayer, where FI is a ferromagnetic insulator with noncollinear magnetisation. We first predict theoretically the formation of a small spectral gap, henceforth called triplet gap, in the LDOS and show how it is related to the creation of equal-spin triplet pairs. We then support the theory through experimental STS studies on the normal side of the trilayer. To our knowledge, such a method for clearly identifying the triplet states spectroscopically has not been reported before, neither theoretically nor experimentally. The fact that equal-spin and mixed-spin states result in distinctly different structures in the LDOS is a novel observation. The triplet gap develops around zero energy, resulting in a symmetric double-peak structure around zero bias voltage in the LDOS. The width of the triplet gap monotonically depends on the ratio of equal-spin
to mixed-spin states in the pairing amplitude. Such a formation of an additional gap within
the superconducting gap, solely depending on the magnetic structure in the proximity of
the superconductor represents a new signature of equal-spin triplet Cooper pairs. Our experimental evidence presented below strongly supports the formation of an equal-spin triplet state, thus making a strong case to pursue superconducting spintronics.

\section*{Results}
\subsection*{Circuit Theory}
\begin{figure}[htp]
  \centering
  \includegraphics{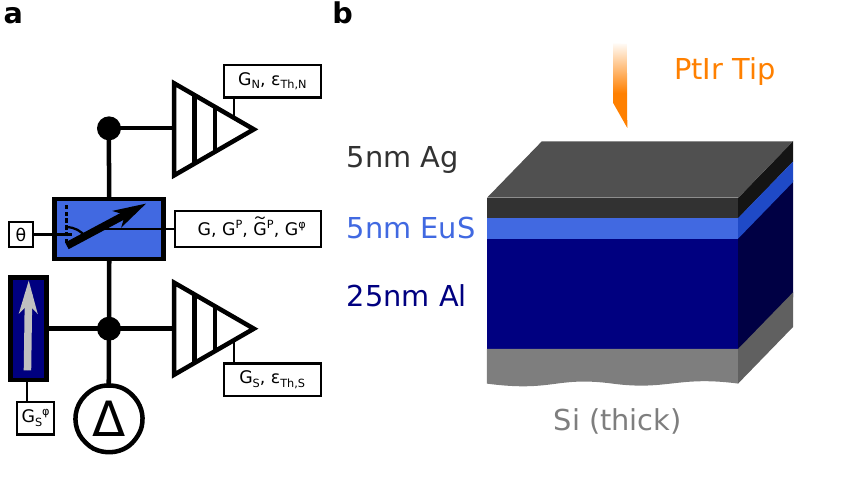}
  \caption{\textbf{Illustration of the theoretical and experimental setup.} \textbf{a}, Circuit diagram representing the theoretical model. A superconducting S-node with
    a $\Delta$-pseudo-terminal is connected to a normal conducting N-node (both with
    conductances $G_{\mathrm{S/N}}$ and Thouless energies $\epsilon_{\mathrm{N/S}}$) via a tunnelling
    connector. The connector has the spin-dependent parameters $\tilde G^P, G^P$ and
    $G^{\phi}$ which depend on the relative magnetisation direction $\theta$ to the spin
    dependent $G^{\phi}$ term of the S-node.  The relative angle $\theta$ between those
    magnetisations is the main free parameter in our fits.
\textbf{b}, Schematic of the tunnel contact. A PtIr tip (normal metal) is
    brought into tunnel contact with a trilayer sample of an EuS layer sandwiched between a normal conducting
    Ag and a superconducting Al film.}
  \label{fig:setup}
\end{figure}

In our theory based on the language of circuit theory \cite{eschrig2015-2, nazarov1994, nazarov1999,huertas2002, braude2007, machon2013, machon2015}, a ferromagnetic insulator separating an s-wave superconductor and a
normal metal can be represented by the circuit diagram depicted in
Fig.~\ref{fig:setup}a. Each conducting layer is represented by one node, characterised
by its conductance $G_{\mathrm{N/S}}$ (the index N/S labels the normal/superconductor side)
and its size-dependent Thouless energy $\epsilon_{\mathrm{Th,N/S}}$. The superconducting layer is specified by
the pseudo terminal characterised by the pair potential $\Delta$, which constitutes a source of coherence
 that has to meet the self-consistency relation \cite{belzig1999}. The
ferromagnetic insulator is described by a connector representing a tunnel barrier with the
conductance $G_{\mathrm T}$. The ferromagnetic nature is accounted for by parameters $G^P$ and $\tilde G^P$ resulting from the spin polarisation of the tunnel probabilities \cite{eschrig2015-2,machon2013},
and the so-called spin-mixing term $G^{\phi}$ \cite{huertas2002}, which all depend on the respective magnetisation directions. We use two
$G^{\phi}$ terms, one in the ferromagnetic connector and one in the superconducting
node, to account for the magnetic texture in the system (which we will discuss later). One advantage of the circuit theory approach is that it starts from very basic concepts and can easily be extended to describe many other realisations of S, F, N heterostructures. Note that in order to realistically represent the investigated circuit with the specific model chosen here, the layer thicknesses must not exceed the coherence length of the superconductor.

\begin{figure}[htp]
  \centering
  \includegraphics{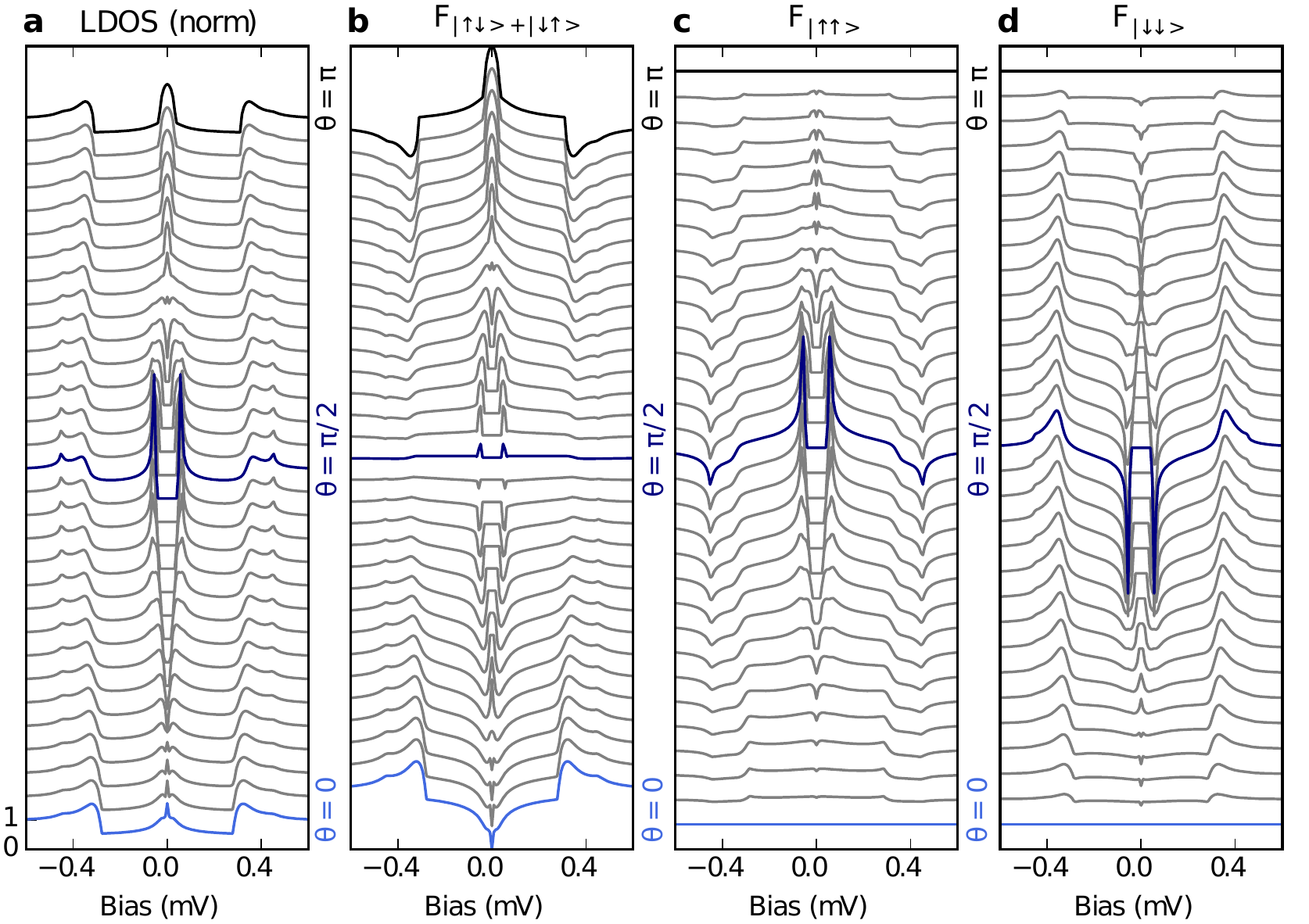}
  \caption{\textbf{Dependence of the superconducting properties on the magnetic configuration.} \textbf{a},
    Evolution of the local density of states (LDOS) as a function of the relative angle $\theta$
    between the magnetisation of the ferromagnetic connector and the $G^{\phi}$ term of
    the S-node, showing zero-bias peaks with varying amplitude for parallel ($\theta = 0$) and
    antiparallel ($\theta = \pi$) alignment and the appearance of the triplet gap around perpendicular
    alignment. \textbf{b}, Pair amplitude of the mixed-spin and (\textbf{c}-\textbf{d}) the
    equal-spin components as a function of $\theta$. The mixed-spin
    component is prominent
    for $\theta \approx 0$ and $\theta \approx \pi$ and almost vanishes around $\theta =
    \pi/2$, while for the equal-spin component it is opposite. All curves have been offset vertically
    for better visibility.
    }
  \label{fig:waterfall}
\end{figure}

The circuit diagram in Fig.~\ref{fig:setup}a represents the discretised version of
the Usadel equation \cite{usadel1970, nazarov1994} and has to be expanded with a
spin-dependent boundary condition, details of which are shown in the Methods section and a full derivation of which can be found in previous publications \cite{eschrig2015-2, machon2013, machon2015}. Solving the Usadel equation (see Eq.~\ref{eq:SN} in the Methods section) allows us to calculate the LDOS in the N-node.
Changing the direction of magnetisation  in the ferromagnetic
connector parametrised by the angle
$\theta$ between the magnetisations of the ferromagnetic insulator's interior and the interface spins, results in an
evolution of features inside the gap shown in
Fig.~\ref{fig:waterfall}a. For $\theta$ close to $0$ and $\pi$, the LDOS shows a
peak at zero bias (corresponding to the Fermi energy). For all other angles, i.e.,
situations where not all magnetisation
directions are collinear, a gap opens symmetrically centred at the Fermi energy. This noncollinear
orientations of magnetisations in the system has been identified as a
mandatory prerequisite of equal-spin triplet pairing \cite{eschrig2008}. Accordingly, the gap is most pronounced at $\theta = \pi/2$, corresponding to a maximal equal-spin pairing. This is visualised in Fig.
\ref{fig:waterfall}c,d, where the equal-spin triplet components of the anomalous Green's
functions of the superconductor are plotted in the $z-$basis projected onto the magnetisation direction
of the S-node $\vec m_{\mathrm{S}}$. It is important to stress that the equal-spin triplet pairing $F_{|\uparrow\uparrow\rangle/|\downarrow\downarrow\rangle}$
has a distinctly different energy dependence than the mixed-spin pairing
$F_{|\uparrow\downarrow\rangle+|\downarrow\uparrow\rangle}$ shown in Fig.~\ref{fig:waterfall}b.
Hence, a full interpretation of the LDOS requires a simultaneous consideration of the energy dependent pair amplitudes.

\subsection*{Scanning tunnelling spectroscopy} 
From the calculated LDOS, the theoretical differential conductance $\mathrm{d}I/\mathrm{d}V$ can be
calculated by including experimental parameters like non-zero temperature and
amplitude of the voltage modulation added to the bias in order to perform lock-in measurements. These calculated curves are compared to $\mathrm{d}I/\mathrm{d}V$ tunnel spectra,
measured in lock-in
technique between a normal metal tip and an Al/EuS/Ag trilayer sample
(see Fig.~\ref{fig:setup}b and Methods section) in a scanning tunnelling microscope (STM) at \SI{290}{\milli \kelvin}, far below the superconducting critical temperature of the Al layer ($T_{\mathrm{c}} = \SI{1.7}{\kelvin}$). As we discuss in Supplementary Note~2, there is strong evidence for the formation of an oxide layer at the interface between Al and EuS. As we will argue below, this oxide layer might be important for the formation of the noncollinear magnetisation arrangement which itself is crucial for the formation of the triplet pairs. The changing direction of magnetisation in the tunnel connector is experimentally realised by exposing the sample to
an external magnetic field parallel to the sample plane.

\subsection*{Spatial dependence of subgap features}
\begin{figure}[htp]
  \centering
  \includegraphics{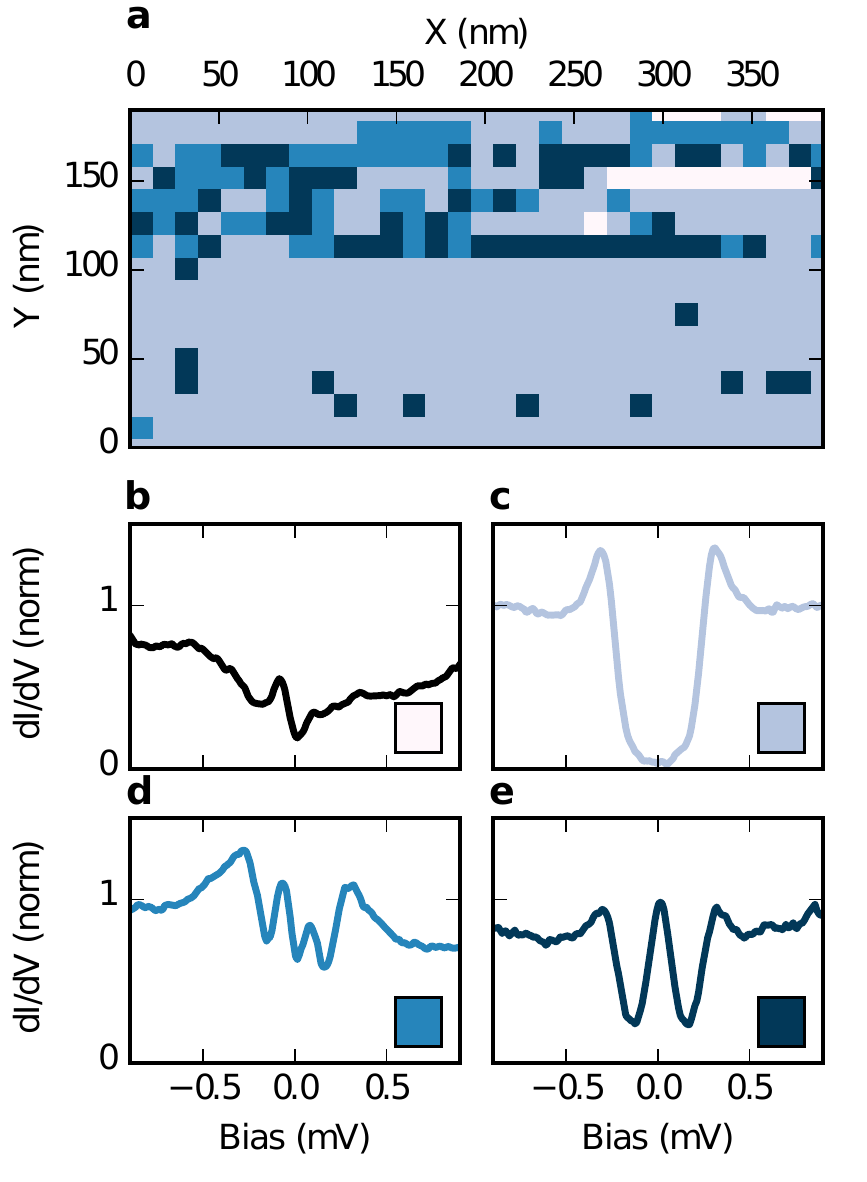}
  \caption{\textbf{Spatial dependence of the differential conductance. a,} Map of locations on the sample surface where different types of spectra shown in \textbf{b-e} have been observed. The different colours in \textbf{a}
    correspond to the different types of spectra observed and match the spectra shown in
    \textbf{b}-\textbf{e}. The shape of the spectra changes on length scales corresponding to the
    grain size of the EuS films (\SI{8}{\nano \meter}-\SI{18}{\nano \meter}) \protect\cite{wolf2013}. Data was recorded on sample EuS-3.}
  \label{fig:spatial}
\end{figure}

In order to characterise the trilayer film sample, we record tunnel spectra (see Supplementary Note~1) by scanning
the tip over the sample with a step size of \SI{12.5}{\nano \meter}. In
Fig.~\ref{fig:spatial} we show a colour-coded map of typical $\mathrm{d}I/\mathrm{d}V$
spectra. All spectra are normalised to the
conductance value $G_{\mathrm{b}}$ far outside the gap. We categorised the spectra in four distinct
groups, each characterised by a specific shape (Fig.~\ref{fig:spatial}b-e). Category (b) corresponds to tip locations where the tunnel contact is too noisy for spectroscopy or where superconductivity is
being suppressed. Spectra of these types are only rarely observed. We attribute these
spectra to surface contamination or defects in the film. Category (c)
shows a spectrum with a hard gap as known from BCS theory, which is the result known for
spin-independent tunnelling and is seen in all Al/Ag bilayer reference samples (see
Supplementary Fig.~1). The gap size $\Delta \approx 250 \pm \SI{20}{\micro \electronvolt}$ is
slightly enhanced with respect to the bulk value of $\Delta_0 = \SI{180}{\micro \electronvolt}$ as usual for
thin Al films \cite{wolf2012}. The most relevant spectra for this work are of
type (d) and (e), in the following referred to as triplet gap and zero-bias peak
spectra, respectively. They show coherence peaks at the gap edge similar to BCS theory,
however, inside the gap the $\mathrm{d}I/\mathrm{d}V$ remains finite everywhere with a minimum value
$\approx 0.25-0.5 G_{\mathrm{b}}$.
The triplet gap spectra (d) feature two peaks located
symmetrically around zero bias, which  are not necessarily symmetric in height.
Category (e) shows a maximum at zero bias the amplitude of which can be larger than
$G_{\mathrm{b}}$. While the majority of the spectra recorded on all measured samples show the BCS-like spectra (c),
some areas show reproducible clusters of zero-bias peaks or triplet-gap features when addressing
the same spot repeatedly. The fact that spectra can swap from triplet gap to zero-bias peak and
to BCS like within one pixel reveals that the phenomenon responsible for the nature of the
features inside the gap can change on short length scales, corresponding to, e.g., the grain size of
the EuS thin film (8-\SI{18}{\nano \meter}) \cite{wolf2013}.

\subsection*{Noncollinear magnetic model} 
To explain our findings we propose the
following model: According to our theoretical studies, the appearance of spectra with
triplet gap features corresponds to areas with at least two magnetisation directions which are
noncollinear. We assume these two distinct magnetic areas are given by the bulk EuS film
on the one hand, and by the interface to the Al layer on the other hand (Fig.
\ref{fig:interface+fieldsweep}a-c), as we will explain in more detail below and in Supplementary Note~3. Second, the ferromagnetic
interface must provide some degree of spin mixing between similarly oriented magnetic domains. These are two non-trivial requirements explaining why only a fraction
of all spectra reveals these features.

\begin{figure}[htp]
  \centering
  \includegraphics{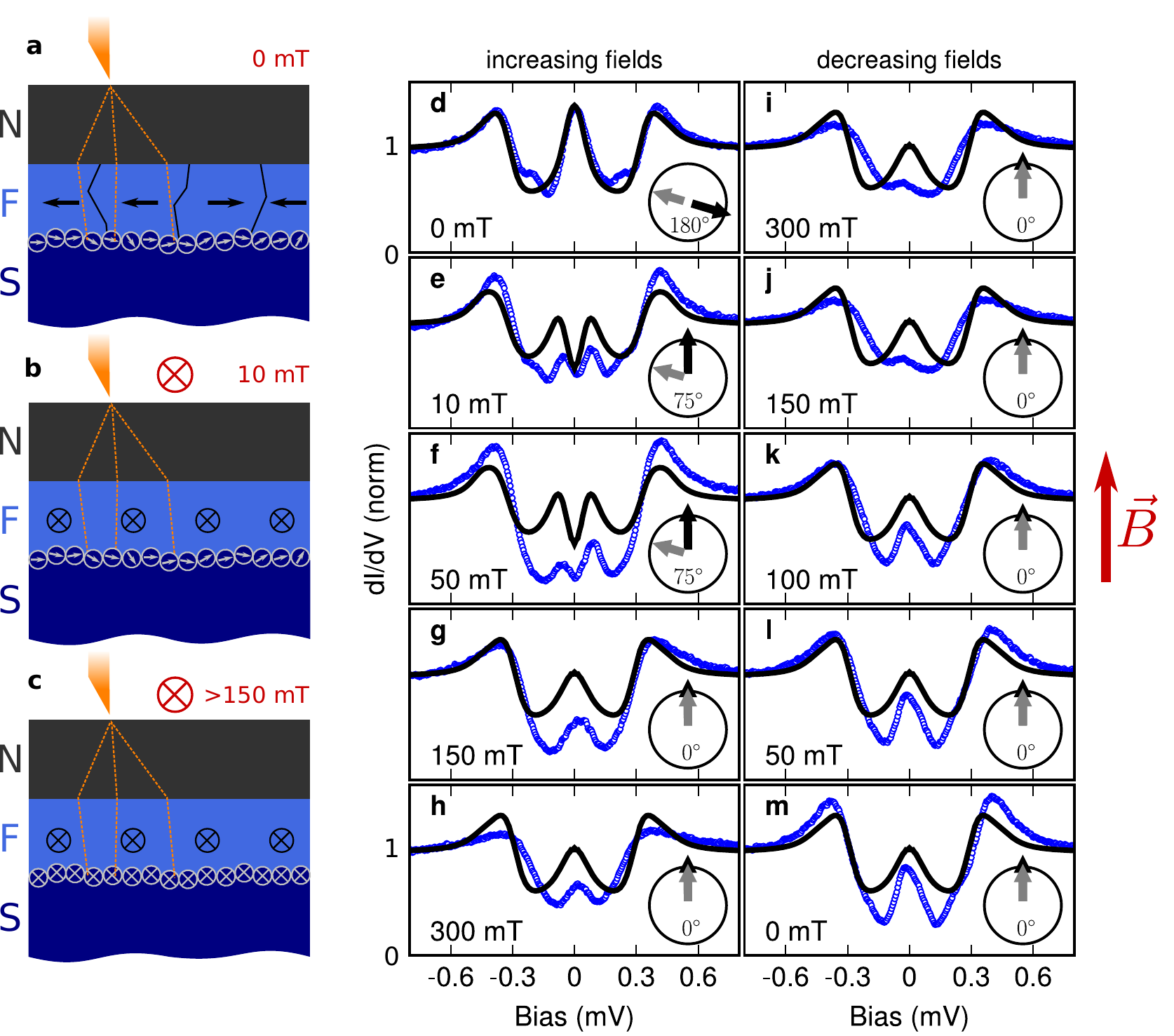}
  \caption{\textbf{Model of the magnetisation behaviour of the EuS layer.} \textbf{a}, The
    sample in the zero-field cooled (ZFC) state consists of magnetically soft domains (black arrows) with an overall
    magnetic moment that is random in direction, and harder interface magnetic moments (grey arrows). \textbf{b}, The internal domains are
    expected to follow the external magnetic field (directed into the plane of projection) more readily, aligning at smaller
    magnetic fields. \textbf{c}, The interface moments follow the applied field only for higher field values. \textbf{d} - \textbf{m}, Experimental
    $\mathrm{d}I/\mathrm{d}V$ spectra (blue circles) recorded for the same tunnel contact in varying magnetic fields
    at \SI{290}{\milli \kelvin} and theoretical spectra (black curves) fitted to the data. A full up and down sweep is performed to show that the observed curves depend on the magnetisation behaviour of the F layer. The black and grey arrows indicate the fitted relative angle between the different magnetisations. Direction of the external field is indicated in red. Data was recorded on sample EuS-1.}
  \label{fig:interface+fieldsweep}
\end{figure}

The dominance of BCS-like spectra indicates that the domain size of the ferromagnetic film is small. 
As wide areas of the EuS film have proven to be nanocrystalline in transmission electron microscopy (TEM)
measurements (see Supplementary Fig.~3), the large number of domains probed simultaneously
results in a minimised average magnetisation. Areas that do show zero-bias peak and triplet-gap
features thus hint at locally enlarged
domain sizes, thereby reducing the effective number of domains being studied at one spot and resulting in a net magnetic moment. Such an extreme sensitivity of the superconducting state to the local domain structure is well known \cite{buzdin2005}. Since no voids or areas with reduced thickness of the EuS film are observed in
the TEM  images, we consider the possibility, that no ferromagnetic material was present in
large parts of the trilayer, to be unlikely.

\subsection*{Magnetic field dependence}
To further verify this model we performed STS measurements with
applied magnetic field, see Fig.~\ref{fig:interface+fieldsweep}. We assume that the
magnetic configuration of a sample in the zero-field cooled (ZFC) state consists of domains with
independent magnetisations pointing in random directions \cite{strambini2017} (black arrows) and of
moments found at the interface of the ferromagnet, that do not necessarily align with the
direction of magnetisation of the underlying bulk domains \cite{bernardo2015} (grey arrows). Once exposed to external magnetic fields, those interface moments follow the direction of the external field only at higher fields compared to the magnetically softer bulk moments. These
interface moments can be magnetically harder due to a local variation of the interface \cite{balkashin2009,miao2014} or surface effects \cite{hellman2017}. The exact nature of the interface moments is irrelevant for the functional principle of the creation of spin-triplet pairs by noncollinear magnetisation. As mentioned above, we observed an oxide layer at the interface between Al and EuS. In Supplementary Note~2 we show that this layer contains grains of EuO. EuO is a ferromagnet with a higher Curie temperature than EuS. Together with the small size of the grains they may be magnetically harder than the bulk EuS film and may help creating locally a noncollinear magnetisation configuration.

We can use the different coercive fields of the oxide interface and the bulk moments of the EuS layer to control the angle
between them by applying an external magnetic field
(Fig.~\ref{fig:interface+fieldsweep}a-c). In order to better understand the relationship
between the features inside the gap and the magnetic properties of the F/S system, we show a
complete up and down sweep of the in-plane magnetic field at a position on the sample,
where a zero-bias peak appears in the ZFC state (for the evolution of a triplet gap feature in magnetic fields see Supplementary Note~4 and Supplementary Fig.~9), and we fit differential conductance
spectra calculated using the model described above to the experimental data
(Fig.~\ref{fig:interface+fieldsweep}d-m). This fitting is done under the constraint that only
the relative magnetisation angle $\theta$ between the oxide interface moments and the bulk moments
can change between different set points of the external field. The parameters characteristic
for the sample geometry ($G_{\mathrm{S/N}}, G^{\phi}_{\mathrm{S/N}}, P$) and the materials used ($T_{\mathrm{c}} = \SI{1.84}{\kelvin}$,
and derived from these self-consistently $\Delta = \SI{280}{\micro \electronvolt}$), were fit to the zero-field spectrum and then kept constant.
We note, that the quality of the fits could be substantially improved by varying these parameters individually for every field independently, which would, however, not be justified by physical arguments. Due to the complexity of the model there are several combinations of spin-dependent parameters (see Supplementary Note~6 and Supplementary Tab.~2) which fit the experimental spectra almost equally well.  However, all these parameter sets correspond to the same evolution of the relevant physical properties (see Supplementary Fig.~11 and Supplementary Fig.~12), i.e., the same magnetisation configurations. The solutions share a strong induced exchange field in the Al (here measured by $G^{\phi}_{\mathrm{S}}$), and a large spin-polarisation $P_n$ of the tunnel current of at least 60~\%. For distinctly smaller values of either $G^{\phi}_{\mathrm{S}}$ or $P_n$ the hallmarks of equal-spin triplets vanish. The material choice is thus crucial for the creation of equal-spin triplet pairs.

The observed field dependence can be consistently interpreted when assuming the tip to be located at an area on the sample,
where the magnetic configuration in the ZFC state is anti-parallel. Anti-parallel configurations between interface moments and bulk magnetisation might be energetically favoured because of their reduced stray-field. Microscopically this could be realised by a magnetically harder layer of interface moments \cite{hellman2017} or by the formation of a ferromagnetic oxidised state of EuS at the EuS/Al interface (see Supplementary Note~2). In the circuit-theory model, this anti-parallel configuration results in a strong
zero-bias peak in the LDOS, which corresponds to the creation of mixed-spin triplet pairs as visible by the peak in the mixed-spin pairing amplitude (Fig.~\ref{fig:waterfall}b). The experimental data follows this prediction closely. As the magnetic field is increased,
the bulk magnetisation readily follows the field direction, and at \SI{10}{\milli \tesla} already the
angle between the interface moments (grey arrows) and the bulk magnetisation (black arrows) is decreased. Fitting this
misorientation to our experimental data results in an angle of 75$^{\circ}$,
which means that the initial bulk magnetisation was rotated in-plane by 105$^{\circ}$ (Fig.~\ref{fig:interface+fieldsweep}e). In our theory model, this rotation from anti-parallel to noncollinear
magnetisations opens up a gap in the LDOS, which directly corresponds to the creation of equal-spin triplet Cooper pairs, as signalled by the increasing equal-spin triplet pairing amplitudes (Fig. \ref{fig:waterfall}c,d). The experimental data clearly reflects this trend. As the magnetic field is further increased, the
oxide interface moments finally also follow  the field direction at around \SI{150}{\milli
  \tesla}. The triplet gap and its confining double peaks disappear to reveal again a zero-bias peak,
  corresponding to collinear magnetisations according to our
fits. This behaviour, the appearance, disappearance, and reappearance of a double peak around zero bias, cannot be explained by a simple Zeeman shift of the LDOS 
 (see Supplementary Note~8 and Supplementary Fig.~11). Further increasing the external field does not substantially change the features inside the gap, but suppression of the superconducting gap by the magnetic field starts at around $|B| \approx \SI{300}{\milli \tesla}$. As expected, here the theoretical model does not describe the experimental spectra any more, since this suppression is most likely due to the onset of  orbital depairing, which would require yet another fit parameter in the theory. As we decrease the external field, no
triplet gap opens up and no double peaks reappear, supporting our assumption that we
started with a magnetic configuration in the anti-parallel state, which we cannot recover
by decreasing the field. However, the zero-bias peak starts reappearing at around
\SI{300}{\milli \tesla} and is fully developed at \SI{100}{\milli \tesla}.
This model for the as-cooled magnetisation configuration also explains why finding such a
transition is so rare – most in-gap features show a much less pronounced field dependence under varying the external magnetic field.

In conclusion, we have shown combined theoretical and experimental evidence that a noncollinear magnetic configuration of the S/FI interface leads to the appearance of a novel type of gap in the superconducting density
of states. This triplet gap is closely related to the creation of equal-spin triplet Cooper pairs because it goes along with a significant increase of the equal-spin triplet pair amplitudes. Zero-bias peaks, contrary to earlier claims, do not hallmark equal-spin triplets, but short-ranged mixed-spin triplets. By selectively tuning the relative magnetisation direction between magnetic moments trapped at the interface and the softer magnetisation of the bulk domains, we are able to significantly influence the LDOS of the system.  Our experiments provide spectroscopic evidence for the superconducting state induced by an FI interlayer with noncollinear magnetic texture. Here, the FI is realized by EuS that not only provides high spin polarisation and effective creation of spin splitting in the superconductor Al, but also builds up an oxide layer between the FI and the superconductor.

Our study also reveals that local variation from a collinear magnetisation arrangement is mandatory to form equal-spin triplet Cooper pairs. The FI thus fulfills several functions: By coupling it to the superconductor it creates spin-triplet correlations, promotes spin-dependent tunnelling of the pair amplitude as well as noncollinear texture on small length scales.
In EuS the texture is given by the grain size and by the formation of an oxide layer between EuS and Al. The texture can be elegantly tuned by an external magnetic field, and thereby the magnitude of the spin-polarised Cooper pairs can be adjusted, thus opening up the possibility for controlling dissipation-less transport
of spin information in spintronics devices.

\section*{Methods}

\subsection*{Circuit Theory.}

The spin-dependent boundary condition of the Usadel equation for an
arbitrary contact depends on the transmission probability $T_n$, polarisation $P_n$ , the
spin mixing angle $\phi_n$ \cite{tokuyasu1988} and the magnetisation axes with unit vector
$\vec m_n$, wherein the index $n$ labels the transport channels.
The number of channels
depends on the size and the shape of the tunnel contact. For simplicity we assume a smooth
contact plane, thus the number of channels $N$ is given as $N = A k_\mathrm{F}^2 / (4 \pi)$, where
$A$ is the tunnelling contact area and $k_{\mathrm{F}}$ the Fermi wave vector.

We work in the tunnel limit ($T_n \ll 1$) and assume small spin mixing ($\phi_n \ll
1$). In this case, the combination of the discrete Usadel equation and the boundary
condition leads to two coupled equations, one for each node
\cite{nazarov1999}, expressing the matrix current conservation
\begin{align}
  \label{eq:SN}
  \check I_{\mathrm S \rightarrow \mathrm N} + \check I_{\mathrm{N}}^L & = 0 = \check I_{\mathrm N \rightarrow \mathrm S} + \check I_{\mathrm {S}}^L\,.
\end{align}
The matrix currents are given by
\begin{align*}
  \check I_{\mathrm N}^L & =  \left[-i \epsilon \frac{G_{\mathrm{N}}}{\epsilon_{\mathrm{Th,N}}}
                 \tau_3,\check G_{\mathrm{N}}\right],\quad
  \check I_{\mathrm S}^L  =   \left[
            x`     \frac{G_{\mathrm S}}{\epsilon_{\mathrm{Th,S}}}\left(-i\epsilon\tau_3+
                 \check \Delta\right) -i G^{\phi}_{\mathrm S} \check \kappa_{\mathrm S}  ,
                 \check G_{\mathrm S}\right]\\
  \check I_{\mathrm S \rightarrow \mathrm N}  &= \frac 12 \left[\frac{G_{ \mathrm T}}{2} \left\{\check G_{ \mathrm S}, \check{\kappa}\right\} \check \kappa + \frac{\tilde G^P}{2}
                                \left[\check G_{\mathrm S}, \check \kappa\right]\check \kappa + G^P\left\{\check G_{\mathrm S}, \check \kappa\right\} -
                                iG^{\phi}\check \kappa, \check G_{\mathrm N}\right] \\
  \check I_{\mathrm N \rightarrow \mathrm S} & = \frac 12 \left [\frac{G_{\mathrm T}}{2} \left\{\check G_{\mathrm N}, \check \kappa \right\}\check \kappa + \frac{\tilde                               G^P}{2}\left[\check G_{\mathrm N}, \check \kappa\right]\check \kappa + G^P \left\{\check G_{\mathrm N}, \check \kappa\right\} - i                                G^{\phi}\check \kappa, \check G_{\mathrm S}\right] \,.
\end{align*}

The equations are closed by demanding the normalisation conditions
$\check G_{\mathrm{S}}^2=1=\check G_{\mathrm{N}}^2$.
The conductances are defined from the experimental interface parameters. The spin-independent parameter
is $G_{\mathrm{T}} = 2 \sum_n T_n$ and the spin-dependent parameters are $\tilde G^P = 2 \sum_n
T_n \sqrt{1-P^2_n}$, $G^P = \sum_n T_n P_n$ and $G^{\phi} = 2\sum_n \delta
\phi_n$. We defined $\check \kappa = \tau_3 \otimes \vec m
\vec \sigma$, with $\tau_i$ and $\sigma_i$ being Pauli matrices in Nambu 
and spin space, respectively. For the theoretical curves shown in Fig.~\ref{fig:waterfall} and Fig.~\ref{fig:interface+fieldsweep}, the following parameters were used:
    $G_{\mathrm{S}}/(G_{\mathrm{T}}\epsilon_{\mathrm{Th,S}})=4.1/(k_{\mathrm{B}}T_{\mathrm{c}})$, $G^\phi_{\mathrm{S}}/G_{\mathrm{T}}=5$,
    $G_{\mathrm{N}}/(G_{\mathrm{T}}\epsilon_{\mathrm{Th,S}})=0.07/(k_{\mathrm{B}}T_{\mathrm{c}})$,$G^\phi/G_{\mathrm{T}}=-0.061$, and $P_n=0.6$.

The directions of the respective
magnetisation directions are denoted  by $\vec m$. The gap matrix is
defined by $\check\Delta=\Delta\tau_1$. Note that the LDOS measured in
the experiment does not depend on the lateral system size and, hence,
equations~\ref{eq:SN} can be normalised e.g. by the conductance of the connector $G_{\mathrm{T}}$, thus
giving a solution independent of the size of the interface area. The leakage parameters can be
further related to sample parameters via
$G_{\mathrm{N/S}}/\epsilon_{\mathrm{Th,N/S}}NG_0 = 8 \pi^2 N_{0,\mathrm{S/N}}
d_{\mathrm{S/N}}/k^2_{\mathrm{F,S/N}} \sim d_{\mathrm{S,N}}/\hbar v_{\mathrm{F,S/N}}$  with the thickness of the
layer $d$, the density of states at the Fermi energy $N_{0,\mathrm{S/N}}$ and
the Fermi wave number/velocity $k_{\mathrm{F,S/N}},v_{\mathrm{F,S/N}}$. We also note,
that the order parameter $\Delta$ has been calculated self-consistently
according to the standard BCS relations.

\subsection*{Sample fabrication and characterisation}

For samples EuS-1 to EuS-4, a \SI{25}{\nano \meter} Al layer, followed by an
EuS film of varying thickness, and a \SI{5}{\nano \meter} capping Ag layer are deposited by e-beam evaporation
on a silicon (111) chip. Sample EuS-5 had a thinner Al layer of $\approx \SI{10}{\nano \meter}$. For this work, five batches of samples were fabricated, with the exact parameters shown in Supplementary Tab.~1.
The substrate and the sample holder are cooled to below \SI{100}{\kelvin} using liquid
nitrogen to grow an Al film of homogeneous thickness with relatively small grain size. At
\SI{25}{\nano \meter}, the surface RMS roughness of Al is smaller than \SI{0.6}{\nano \meter},
thus forming a smooth surface for the subsequent EuS and Ag layers. The EuS layer is evaporated onto the substrate at higher temperatures 
to provide a compound with a Curie temperature (see Supplementary Fig.~5) close to the value reported in the literature ($T_{\mathrm{Curie}}= \SI{16.7}{\kelvin}$ \cite{hao1990}). At room
temperature, EuS is semiconducting with an indirect energy gap,
the conduction band minimum at \SI{300}{\kelvin} is \SI{1.64}{\electronvolt}
\cite{li2013}. At cryogenic temperatures, EuS is insulating with resistivity of around
$\rho = \SI{e4}{\ohm \centi \meter}$ for high-quality single crystals
\cite{yang2014}. 
Disorder at the atomic level and unintentional doping can lower this
resistivity by several orders of magnitude and at the same time increase the Curie
temperature (as observed by SQUID magnetometry measurements in Supplementary Fig.~6)
due to interactions between charge carriers and the Eu$^{2+}$ ions
\cite{yang2014}. Because of the presence of an additional insulating oxide film
under the EuS layer (see Supplementary Fig.~3 and Supplementary Fig.~4), we assume the ferromagnetic insulator layer to have very low
conductance, i.e. enabling only tunnel transport. The surface RMS roughness of this oxide layer is $\approx \SI{0.52}{\nano \meter}$ for line profiles recorded on TEM lamellas.

The deposition rate for the EuS film is \SI{0.1}{\nano \meter \per \second}, and the final thickness of the
film is varied from sample to sample (each chip holding samples with four different
thicknesses) using a sample holder with a movable shutter that allows parts of the chip to
be covered during evaporation. The surface RMS roughness of the EuS layer is $\approx \SI{0.59}{\nano \meter}$ for line profiles recorded on TEM lamellas. However, during STS measurements we often observe variations of
the spectra taken on different locations on the same sample to be more pronounced
than the differences between the various film thicknesses. This supports our theory that
the EuS film has only vanishing magnetic influence on a proximitised superconductor when
nanocrystalline, while regions with more uniformly magnetised domain clusters or locally
enlarged domains can strongly influence the superconducting state. Thus, the
lateral extent of the magnetically ordered region seems to play a much larger role than the film thickness.

Some of our EuS films were found to be conductive,
allowing for a proximity coupling between Al and Ag that is much stronger than
anticipated. These samples generally displayed triplet gap features only for very few locations
on the film, or often not at all. Al is a suitable superconductor for this study because of its long
spin relaxation time \cite{meservey1994}. Thus, electrons that have experienced
spin-mixing at the EuS interface can carry that information far into the
superconductor. The bulk critical field of Al ($\mu_0 H_{\mathrm{c}}$ = \SI{10}{\milli \tesla}) is
significantly increased in thin films (see Supplementary Fig.~2). For in-plane fields at \SI{25}{\nano \meter}
thickness it is around \SI{800}{\milli \tesla} \cite{meservey1994}. In order to protect
the EuS layer from contamination and to have a normal metal layer providing
clean metallic surface necessary to perform STM and STS, Ag is an ideal choice for the top layer, since
the proximity effect of diffusive Al/Ag bilayers is well understood and is reliably
described by the Usadel equation \cite{wolz2011}. The surface RMS roughness of the entire Ag-capped multilayer film is smaller than $\SI{0.7}{\nano \meter}$. The Al(\SI{25}{\nano
  \meter})/EuS(\SI{5}{\nano \meter})/Ag(\SI{5}{\nano \meter}) (see Fig.~\ref{fig:setup}a)
multilayers have a critical temperature of $T_{\mathrm{c}} \approx \SI{1.7}{\kelvin}$ (see Supplementary Fig.~7), similar to the
critical temperature of the Al(\SI{25}{\nano \meter})/Ag(\SI{5}{\nano \meter}) reference
sample. 

High-resolution transmission electron microscopy (TEM) images (see Supplementary Fig.~3) show all films (including the oxide layer between the Al and EuS films) to be nanocrystalline with visible lattice planes within most grains. SQUID magnetometry (see Supplementary Fig.~5) and further subgap features (see Supplementary Note~7 and Supplementary Fig.~13) are shown in the supplementary information.

\subsection*{Scanning tunnelling microscopy}
Scanning tunnelling spectroscopy with an IrPt tip is performed in a $^{3}$He cryostat at
\SI{290}{\milli \kelvin}, which results in differential conductance $\mathrm{d}I/\mathrm{d}V$ vs. voltage
$V$ spectra of the types shown throughout this work. All spectra are normalised to the
conductance value far outside the gap. Spectra were recorded in lock-in technique across a \SI{10}{\mega \ohm}
tunnelling gap, the STM set point was set to a tunnelling current of \SI{400}{\pico
  \ampere} at \SI{4}{\milli \volt} tip voltage and the feedback loop was then stopped before voltage sweeps. The STM used to conduct this study was home built in Konstanz \cite{debuschewitz2008} and optimised for high energy resolution spectroscopy at low temperatures \cite{debuschewitz2007}. The STM controller used was a commercially available SPM1000 by RHK with a proprietary current amplifier (IVP-300) for STM studies. The amplitude of the AC voltage modulation added to the tip bias voltage was set between \SI{7}{\micro \volt} and \SI{20}{\micro \volt} at a frequency of \SI{733}{\hertz}.

\section*{Acknowledgements}
We are grateful to A. di Bernardo, F. Giazotto, H. v. Löhneysen, P. Leiderer, J. Moodera, O. Millo, J.W.A. Robinson,  R. Schneider and C. Strunk for fruitful discussion. We thank M. Wolf for preparing early sample batches, M. Krumova and R. Schneider %from the Laboratory for Electron Microscopy (LEM) at KIT 
for providing the TEM data, S. Andreev and G. Fischer %from the Physikalisches Institut (PHI) at KIT 
for carrying out the SQUID magnetometry measurements.

This work was partially funded by a scholarship according to the
Landesgraduiertenförderungsgesetz, by the Baden-Württemberg foundation in the framework of
research network Functional Nanostructures, by the Deutsche Forschungsgemeinschaft (DFG)
through SPP 1538 Spincaloric Transport and a Deutsch-Iraelisches Projekt, as well as by
the Leverhulme Trust.

\section*{Author Contributions}

E.S. and W.B. conceived and developed the project. C.S. fabricated the samples, C.S. and D.B. performed sample characterisation studies. S.D. and M.W. performed the
tunnel experiments. P.M. and W.B. analysed and described the theoretical model and carried out the numeric implementation of the theory. S.D. analysed the data and performed the fitting procedure. S.D., P.M., C.S., W.B., and E.S. wrote the paper, all authors discussed the results and the manuscript.

\section*{Competing Interests}

The authors declare no competing interests.

\section*{Additional Information}

Correspondence and requests for materials
should be addressed to E.S. (elke.scheer@uni-konstanz.de) for the experimental part, and W.B. (wolfgang.belzig@uni-konstanz.de) for the theoretical work. 

\section*{Data Availability}
The datasets generated during and analysed during the current study are available from the corresponding authors on reasonable request. The compiled custom computer code applied during the current study is available from the corresponding author W.B. on reasonable request.

\end{document}

\maketitle
\clearpage

\begin{figure}[htp]
  \centering
\includegraphics{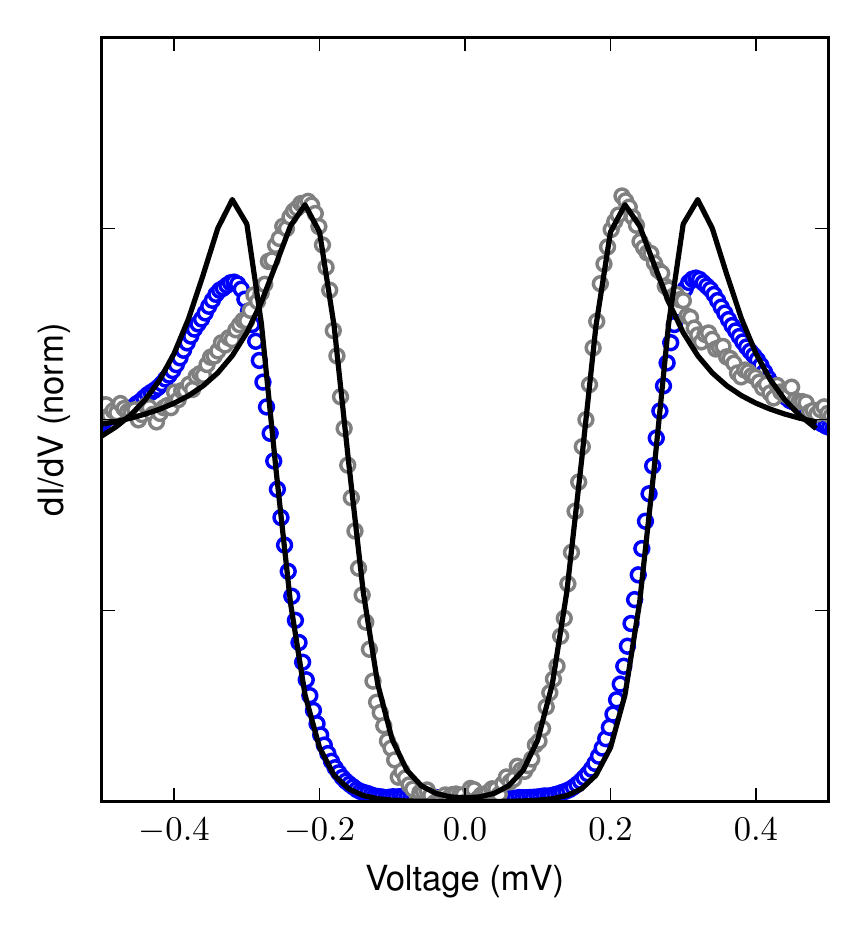}
\caption{\textbf{Reference $\mathrm{d}I/\mathrm{d}V$ measurements at \SI{290}{\milli \kelvin} on a \SI{25}{\nano \meter} (blue) and a
  \SI{240}{\nano \meter} (grey) Al film}. Both films were capped with \SI{5}{\nano \meter} Ag. Fitting
  $\mathrm{d}I/\mathrm{d}V$ spectra calculated based on the Usadel equation (black) results in $\Delta =
  \SI{285}{\micro \electronvolt}$ and $\Delta = \SI{185}{\micro \electronvolt}$,
  respectively, and an effective electron temperature of \SI{295}{\milli \kelvin}. Data was recorded on sample EuS-3.}
  \label{fig:reference}
\end{figure}

\begin{figure}[htp]
  \centering
\includegraphics{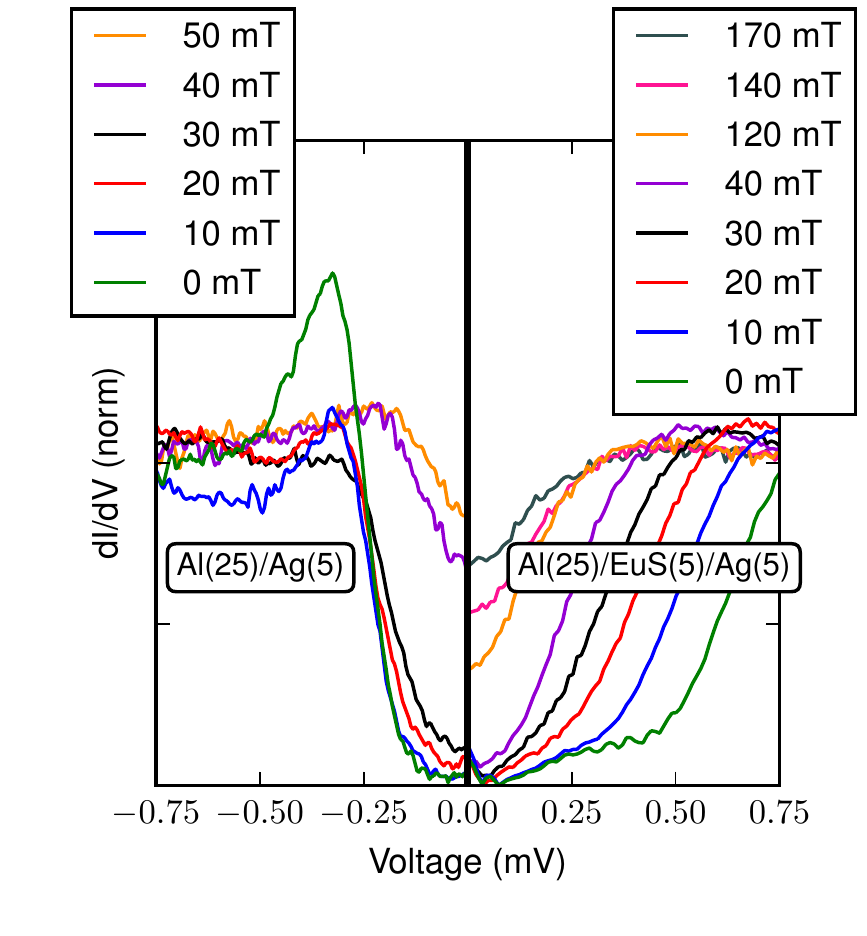}
\caption{\textbf{Comparison between Al/EuS/Ag and Al/Ag reference sample.}
Magnetic field dependent spectra of a Al(25)/EuS(5)/Ag(5) sample 
  show a higher critical field than the Al(25)/Ag(5) reference sample. Data was recorded on sample EuS-2.}
  \label{fig:5nm-vs-ref}
\end{figure}

\begin{figure}[htp]
  \centering
\includegraphics{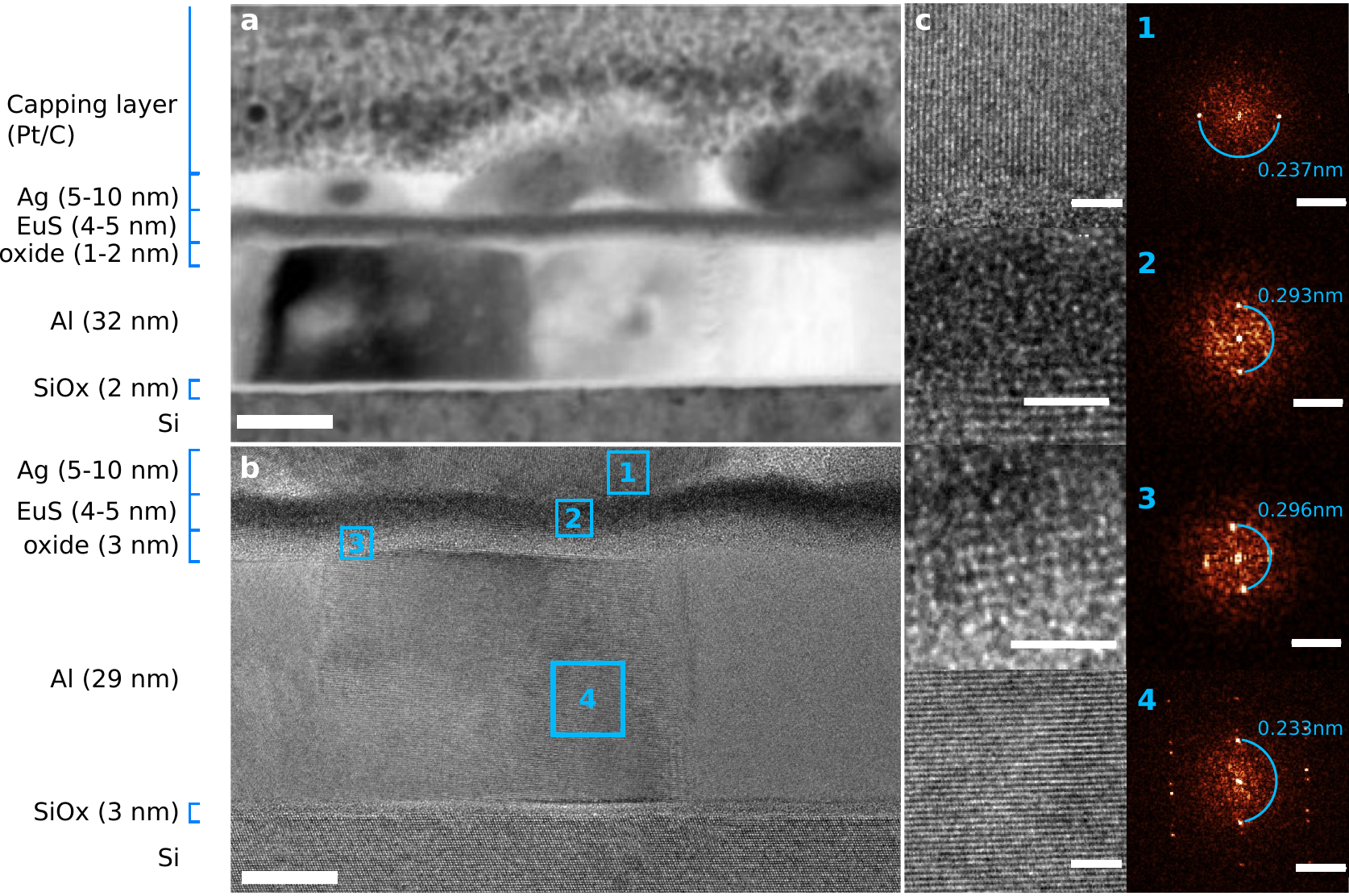}
\caption{\textbf{Transmission electron microscopy (TEM) images of a Al(25nm)/EuS(5nm)/Ag(5nm) lamella. a,} The scanning transmission electron micrograph shows that growth of the Al and EuS layers are uniform across the sample, with slightly thicker Ag and Al films expected. A bright oxide layer is visible between the Al and EuS films (the scalebar has a length of \SI{10}{\nano \meter}). \textbf{b,} The high resolution micrograph shows all films (including the oxide layer) to be nanocrystalline with visible lattice planes for most grains. The oxide layer varies in thickness and shows a gradual transition to the EuS layer and a sharp border to the Al film (the scalebar has a length of \SI{10}{\nano \meter}). \textbf{c}, The reciprocal space information (Fast Fourier Transform) shows lattice planes for grains of all films with the expected reflections from the (111) planes for Ag and Al and reflections from the (200) planes for EuS. The reflections in the oxide layer match various Eu oxides, and none of the common reflections of Al oxides (the scalebars have a length of \SI{2}{\nano \meter} for the zoomed topographies and \SI{5}{\per \nano \meter} for the reciprocal space information images). Images were recorded on sample EuS-4.}
\label{fig:tem}
\end{figure}

\begin{figure}[htp]
  \centering
\includegraphics{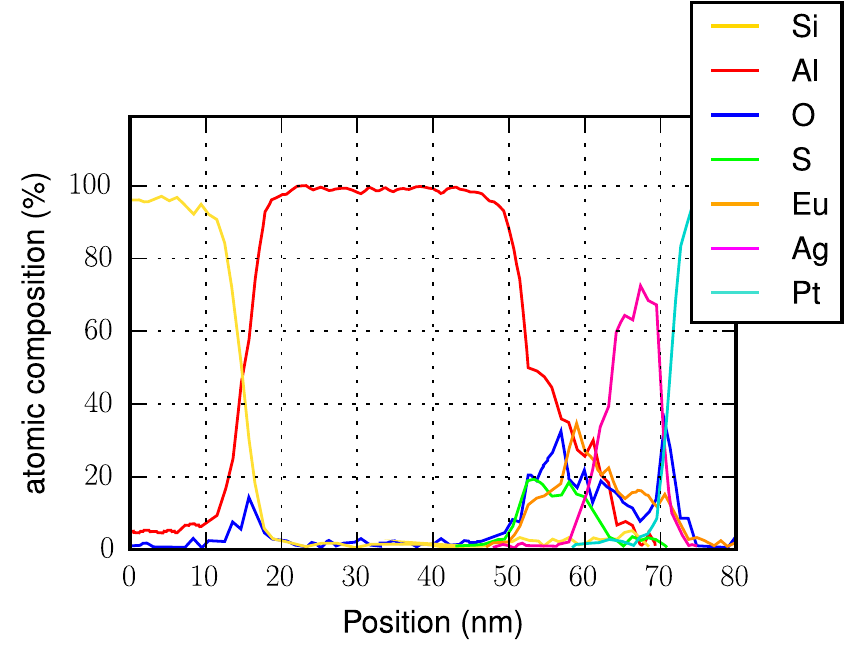}
\caption{\textbf{Elemental analysis of one Al/EuS/Ag sample by energy dispersive X-ray (EDX)
  spectroscopy.} The measurement was performed 2.5 months after the sample fabrication. The precence of O between the Al and EuS films is visible. Data was recorded on sample EuS-2.}
  \label{fig:edx}
\end{figure}

\begin{figure}[htp]
  \centering
\includegraphics{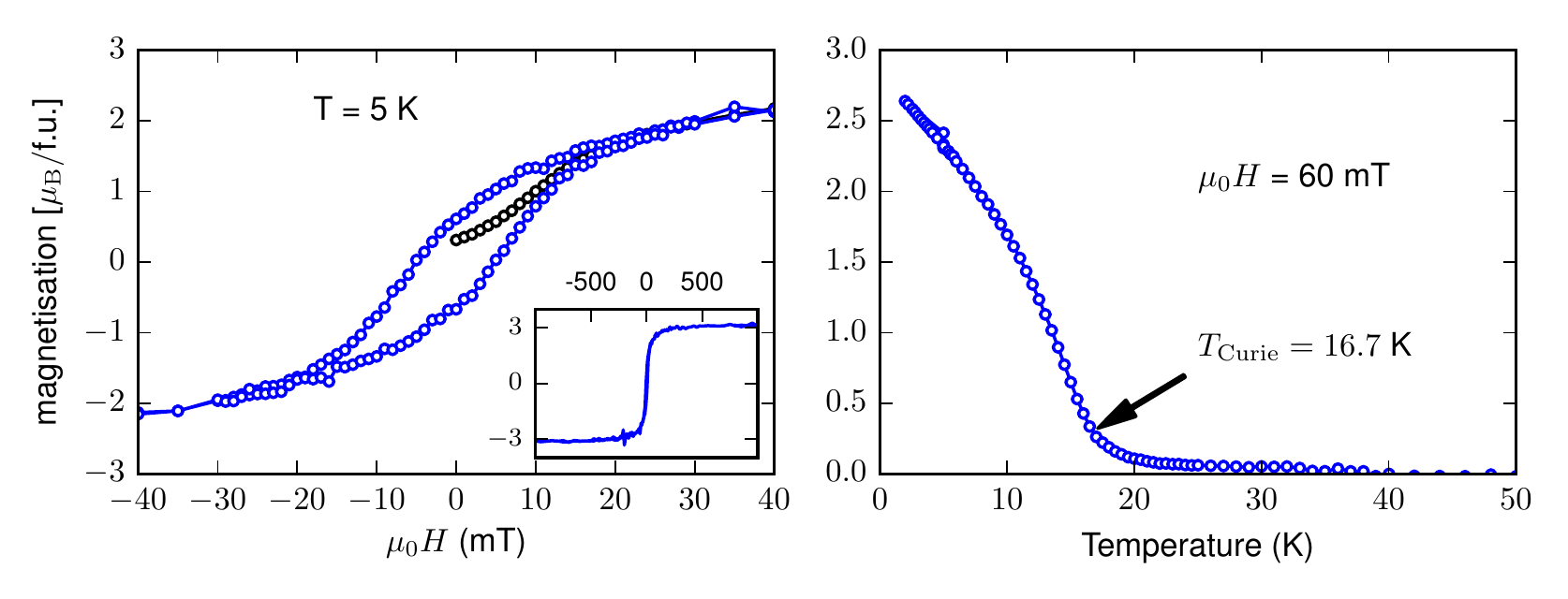}
\caption{\textbf{SQUID magnetometry measurement in parallel field on one of the Al/EuS/Ag samples (no STS measurements performed on this sample).} The sample is
  magnetically soft with a coercive field of $\mu_0 H_{\mathrm c} \approx \SI{5}{\milli \tesla}$ and a Curie temperature of $T_{\textrm{Curie}}
  \approx \SI{16.7}{\kelvin}$, close to the Curie temperature of the bulk material. When zero-field cooled, the film shows almost no spontaneous magnetisation (black curve). The diamagnetic moment of the Si substrate was subtracted from the magnetisation curve. Data was recorded on sample EuS-1.}
  \label{fig:squid-kn0}
\end{figure}

\begin{figure}[htp]
  \centering
\includegraphics{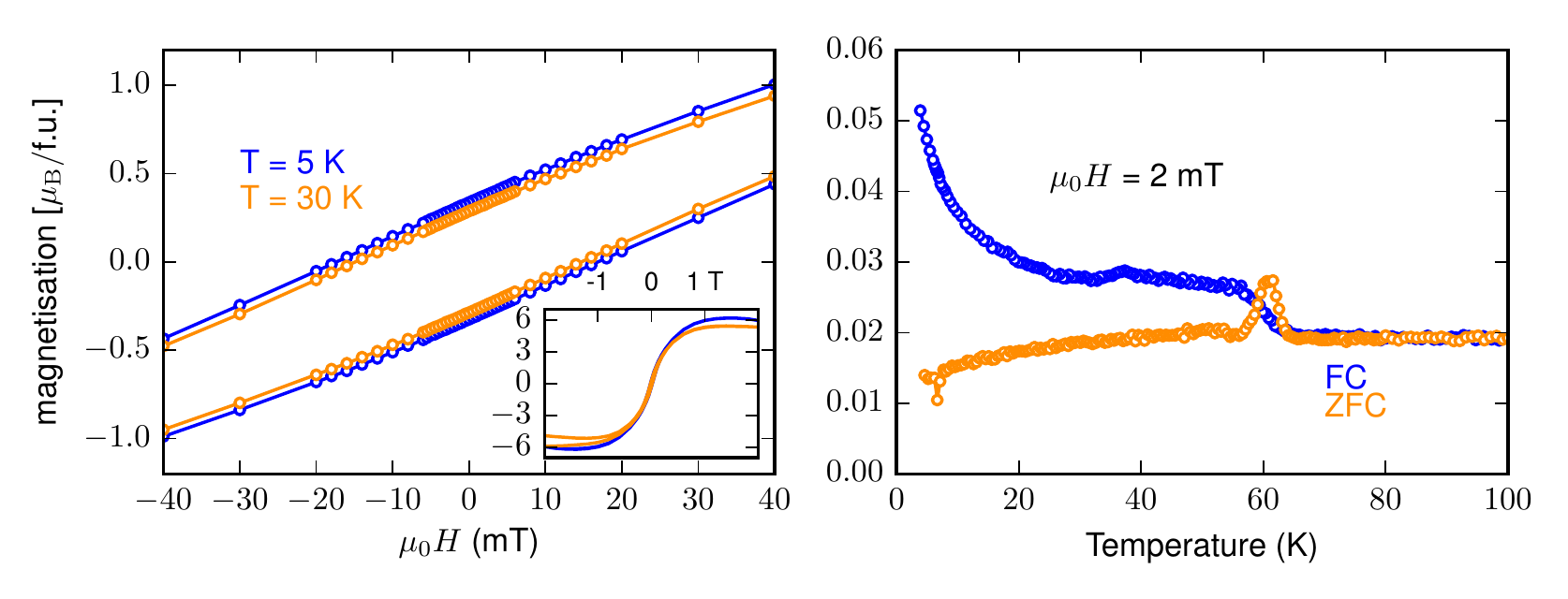}
\caption{\textbf{SQUID magnetometry in out-of-plane field on Al/EuS/Ag sample EuS-2.} The sample is
  magnetically soft both at \SI{5}{\kelvin} and at \SI{30}{\kelvin} with a coercive field of \SI{15}{\milli \tesla} and two different Curie temperatures of $T_{\textrm{Curie}_1}
  \approx \SI{17}{\kelvin}$ and $T_{\textrm{Curie}_2}
  \approx \SI{63}{\kelvin}$, which are close to the exepected Curie temperatures of bulk EuS and EuO, respectively. When zero-field cooled, the film shows a distinct peak in the $M(T)$ curve (orange) around \SI{63}{\kelvin}. Data was recorded on sample EuS-2.}
  \label{fig:squid-kn1}
\end{figure}

\begin{figure}[htp]
  \centering
\includegraphics{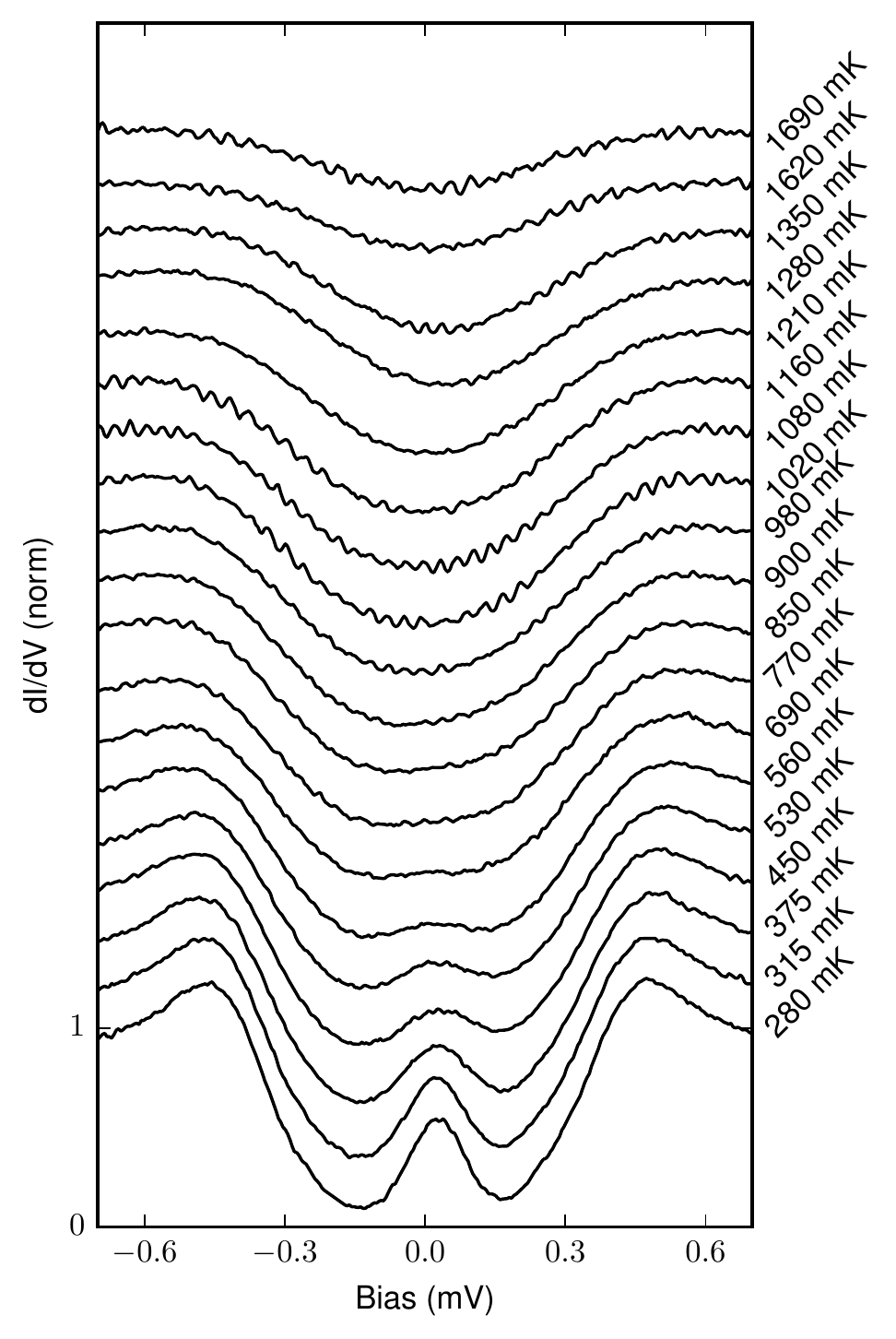}
\caption{\textbf{Temperature dependence of the superconducting spectra of an Al(\SI{25}{\nano
    \meter})/EuS(\SI{5}{\nano\meter})/Ag(\SI{5}{\nano \meter}) sample.} The gap includes a zero-bias peak vanishing around \SI{530}{\milli \kelvin} and shows a
  second-order phase transition with $T_{\mathrm{c}} \approx \SI{1.7}{\kelvin}$. All spectra are offset vertically for better visibility. Data was recorded on sample EuS-1.}
  \label{fig:temp}
\end{figure}

\begin{figure}[htp]
  \centering
\includegraphics{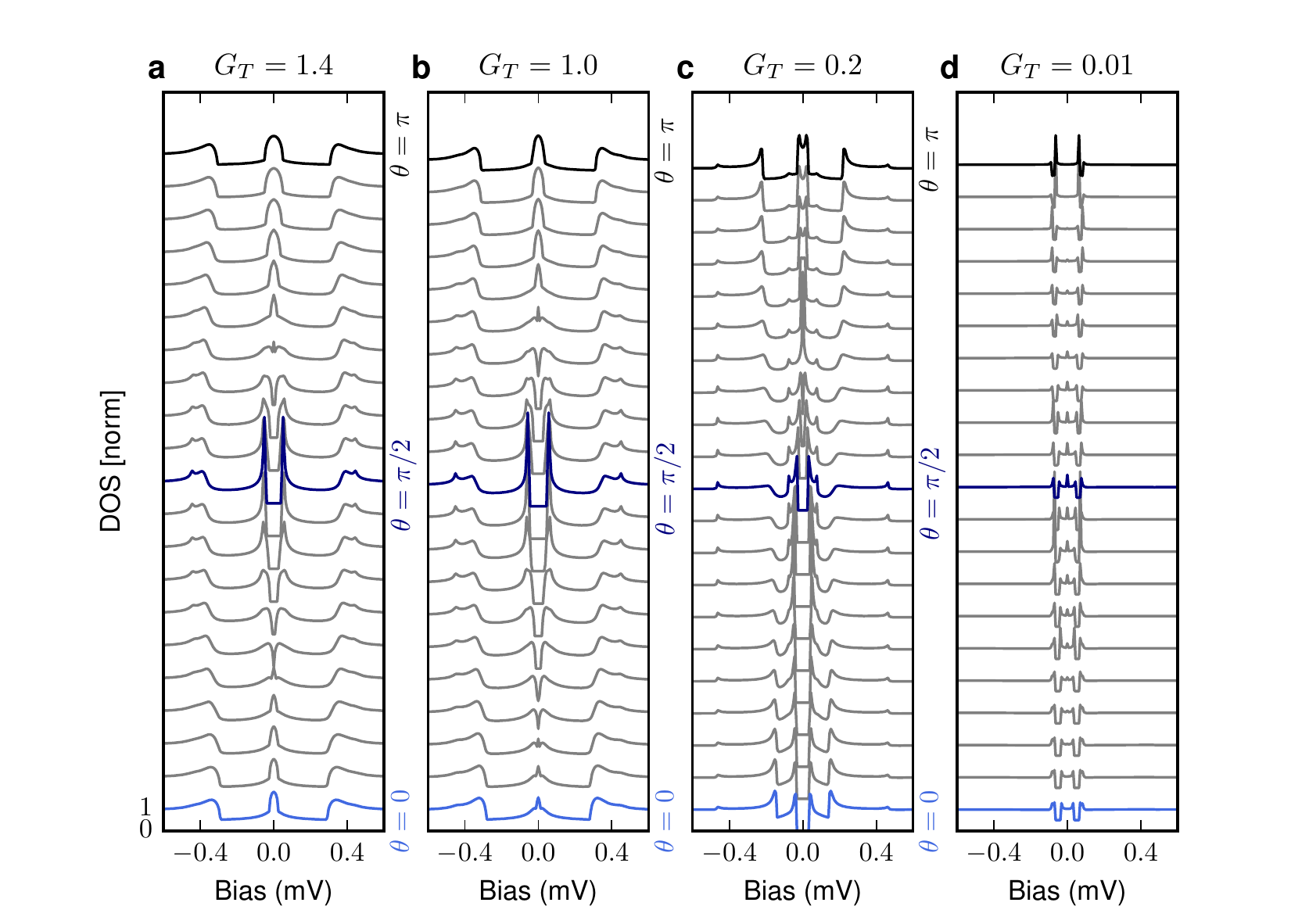}
\caption{\textbf{Calculated LDOS at the normal node for different values of the tunnelling conductance $G_{\mathrm T}$.} For low values of $G_{\mathrm T}$, i.e., a well-insulating tunnel barrier, the subgap features get progressively suppressed, while for high values of $G_{\mathrm T}$ the subgap features grow in amplitude.}
  \label{fig:gt-waterfall}
\end{figure}

\begin{figure}[htp]
  \centering
  \includegraphics[scale=0.8]{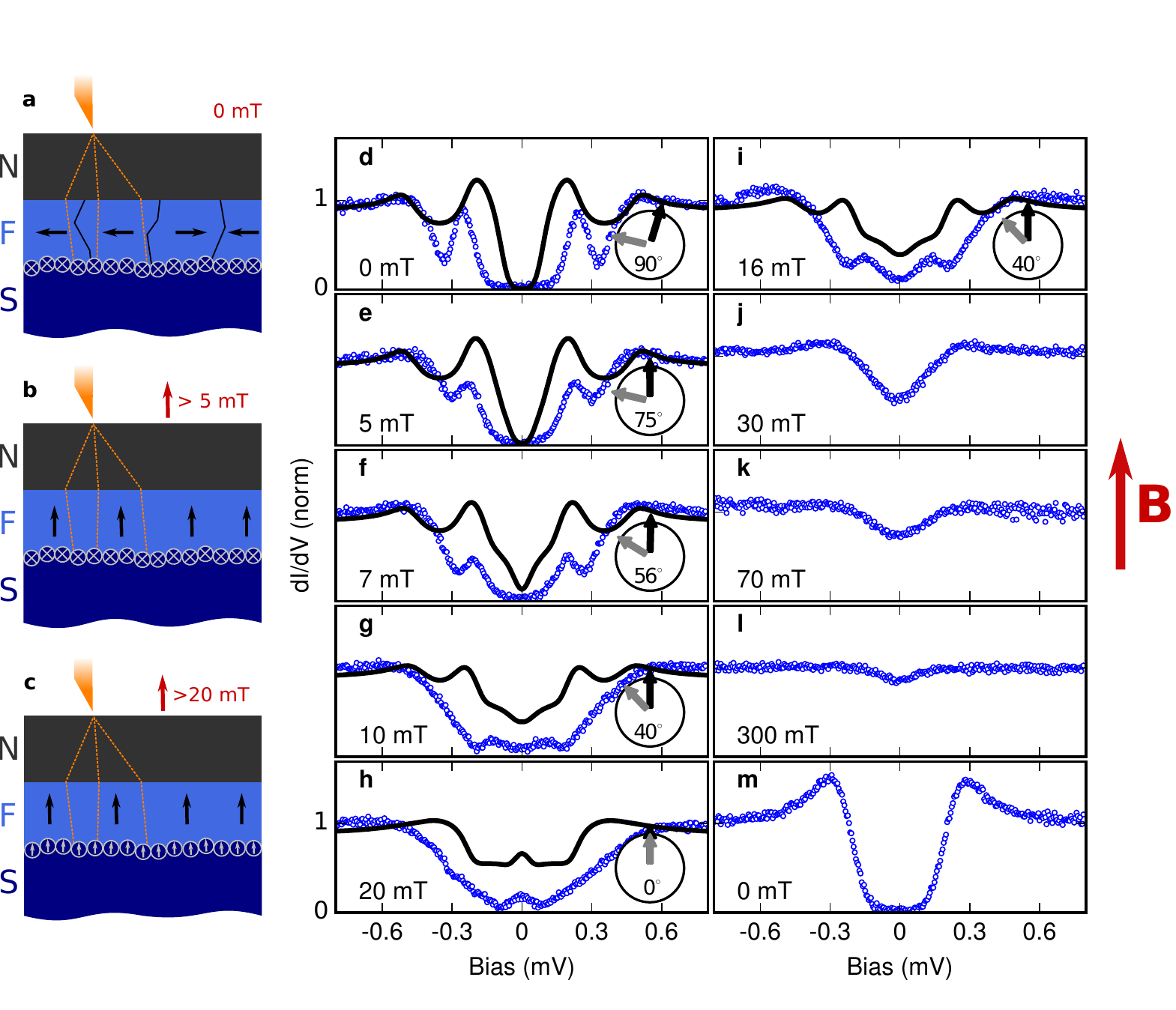}
  \caption[Model of the magnetisation behaviour of the EuS layer for out-of-plane
  fields]{\textbf{Model of the magnetisation behaviour of the EuS layer for out-of-plane fields.} \textbf{a}, The
 sample in the as-cooled state consists of magnetically soft domains with an overall
 magnetic moment that is random in direction, and interface moments pinned by
 impurities which show higher coercive fields. \textbf{b}, The internal domains are
 expected to follow the external magnetic field more readily, aligning at smaller
 magnetic fields. The interface moments have started to follow the field, but are not
 aligned yet. \textbf{c}, The interface moments only follow the external 
 magnetisation for higher fields. \textbf{d} - \textbf{h}, Experimental
 $\mathrm{d}I/\mathrm{d}V$ spectra (blue) recorded for the same tunnel contact in varying magnetic fields
 at \SI{290}{\milli \kelvin}. A magnetic field sweep is performed to show that the observed
 curves depend on the magnetisation behaviour of the F layer. The black lines are the
 calculated differential conductances according to our model. The arrows
 indicate the fitted relative angle between the different magnetisations. Data was recorded on sample EuS-2.}
  \label{fig:si-double-peak-fieldsweep}
\end{figure}

\begin{figure}[htp]
  \centering
  \includegraphics[scale=0.8]{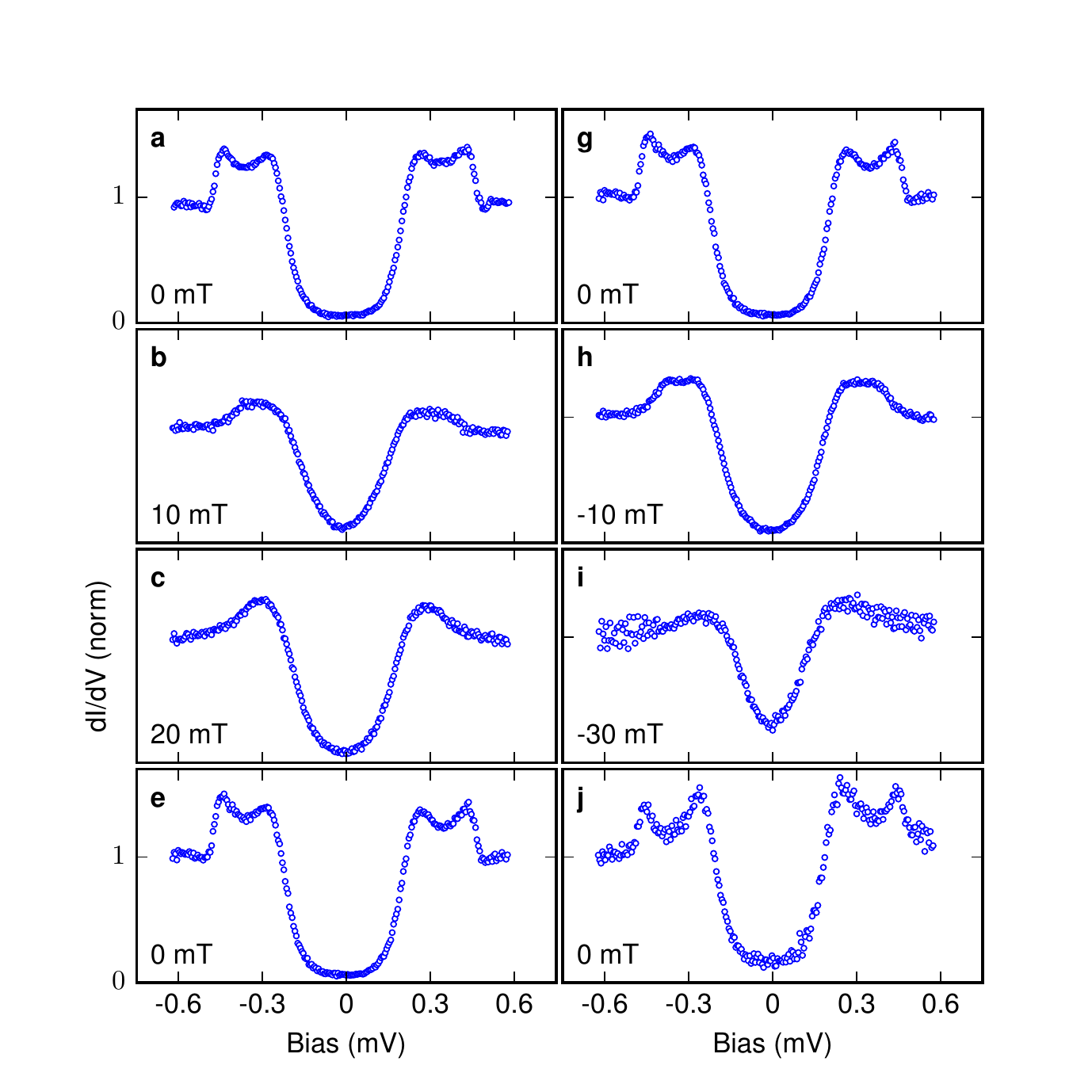}
  \caption{\textbf{Field dependence of a wide double-peak spectrum, exposed to an out-of-plane magnetic field in both directions.} No hysteresis is detected upon field reversal, and the double peak spectrum returns once the field is turned off. Spectra were recorded at 290mK on an Al(25nm)/EuS(5nm)/Ag(5nm) trilayer. Data was recorded on sample EuS-2.}
  \label{fig:hysteresis}
\end{figure}

\begin{figure}[htp]
  \centering
  \includegraphics{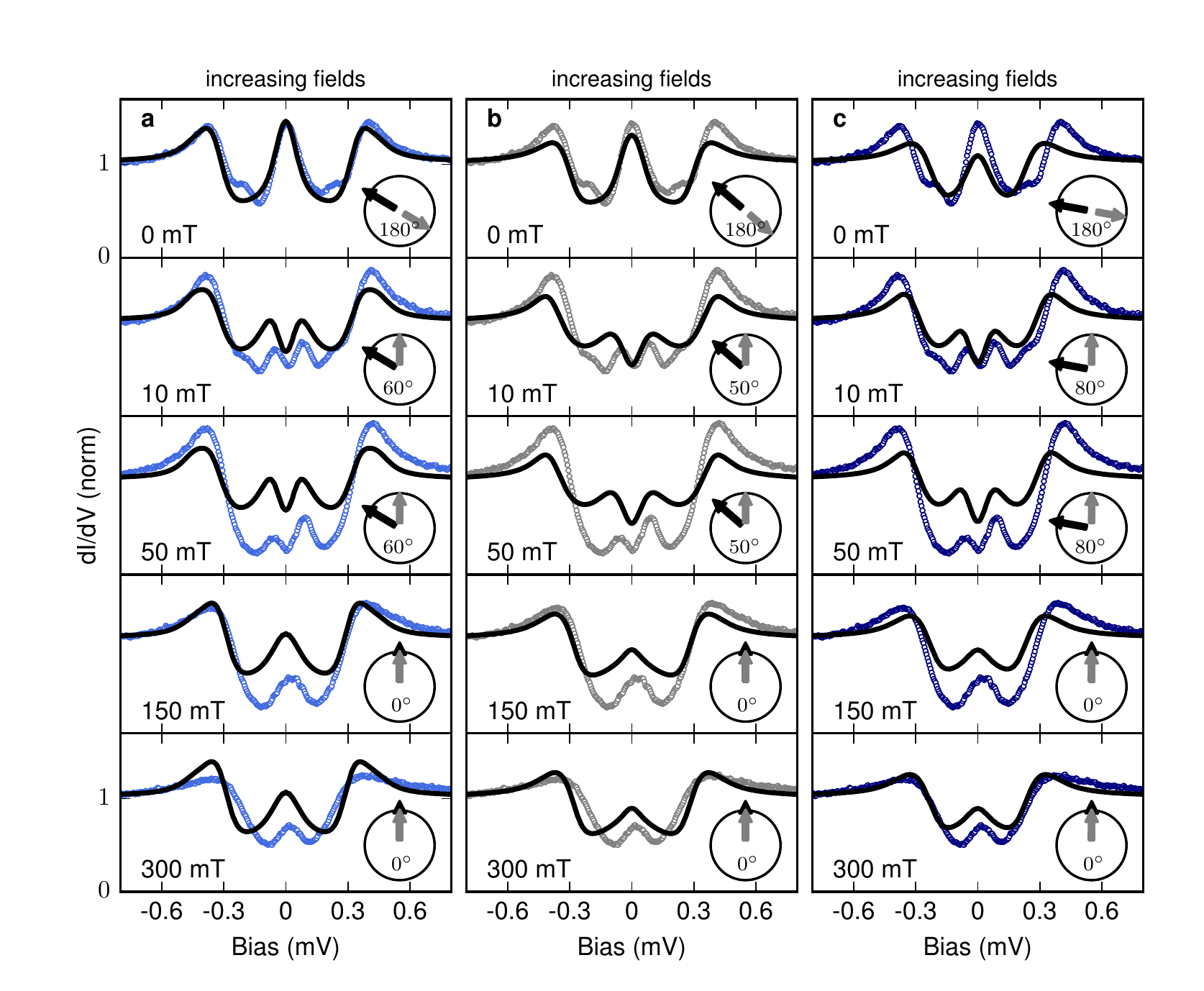}
  \caption{\textbf{Different parameter combinations for fitting the experimental spectra of Fig. 4. a,} 
shows the theoretical curves displayed in the main article, \textbf{b-c} are show alternative parameter combinations.
The numerical parameters for the theoretical curves can be found in Supplementary Tab.~\ref{tab:parameters}, experimental spectra are
identical to the ones shown in the main article (Fig.~4. For the parameters used, 
see Supplementary Tab.~\ref{tab:parameters}, for LDOS and triplet pairing amplitudes see Supplementary Fig.~\ref{fig:si-waterfall}.}
  \label{fig:si-fieldsweep}
\end{figure}

\begin{figure}[htp]
  \centering
  \includegraphics{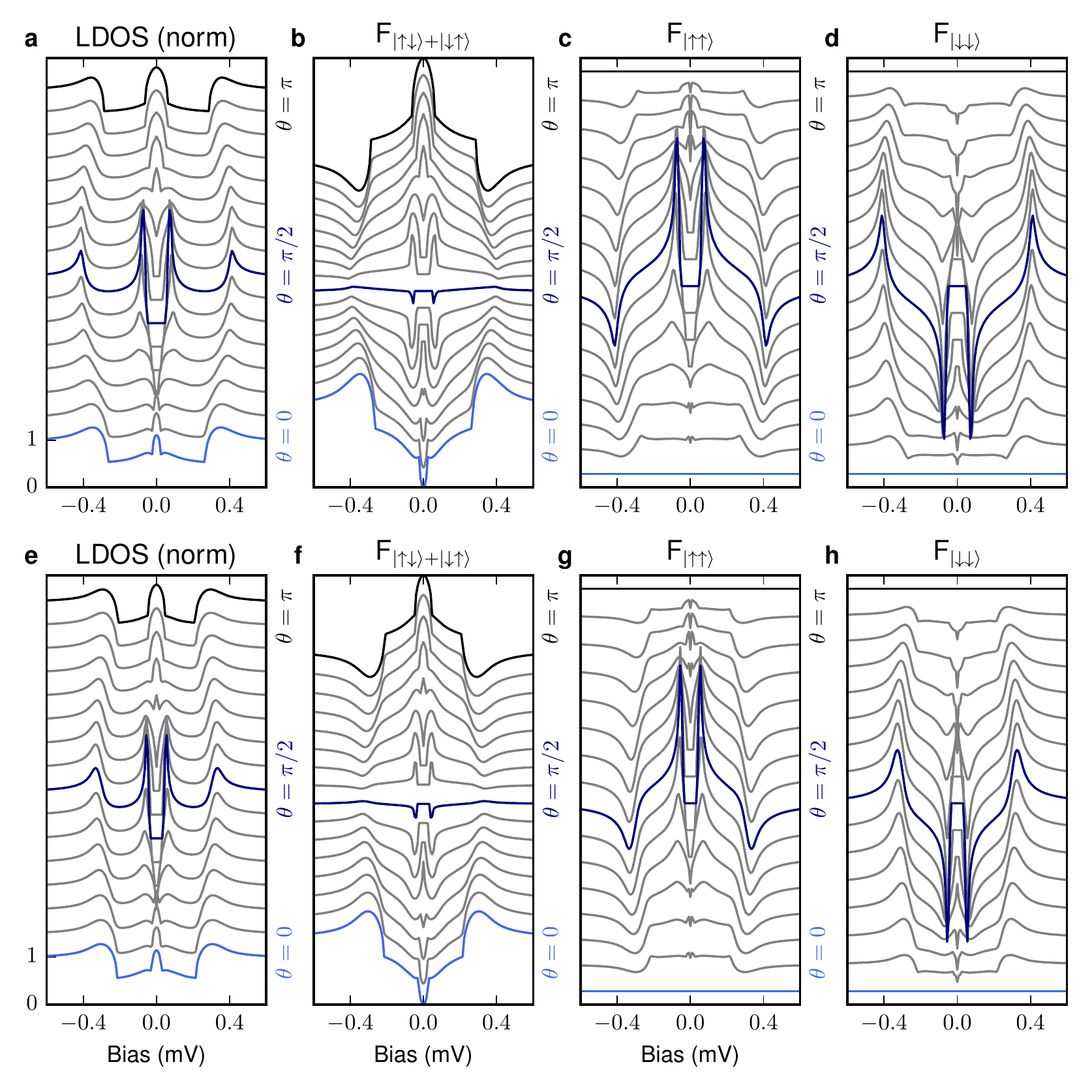}
\caption{\textbf{Dependence of the superconducting properties of the S/FI/N system on the magnetic configuration. a-d,} LDOS and triplet 
pairing amplitudes for the $\mathrm{d}I/\mathrm{d}V$ spectra shown in~\ref{fig:si-fieldsweep}b, \textbf{e-h} the same for the spectra shown in~\ref{fig:si-fieldsweep}c}
  \label{fig:si-waterfall}
\end{figure}

\begin{figure}[htp]
  \centering
  \includegraphics{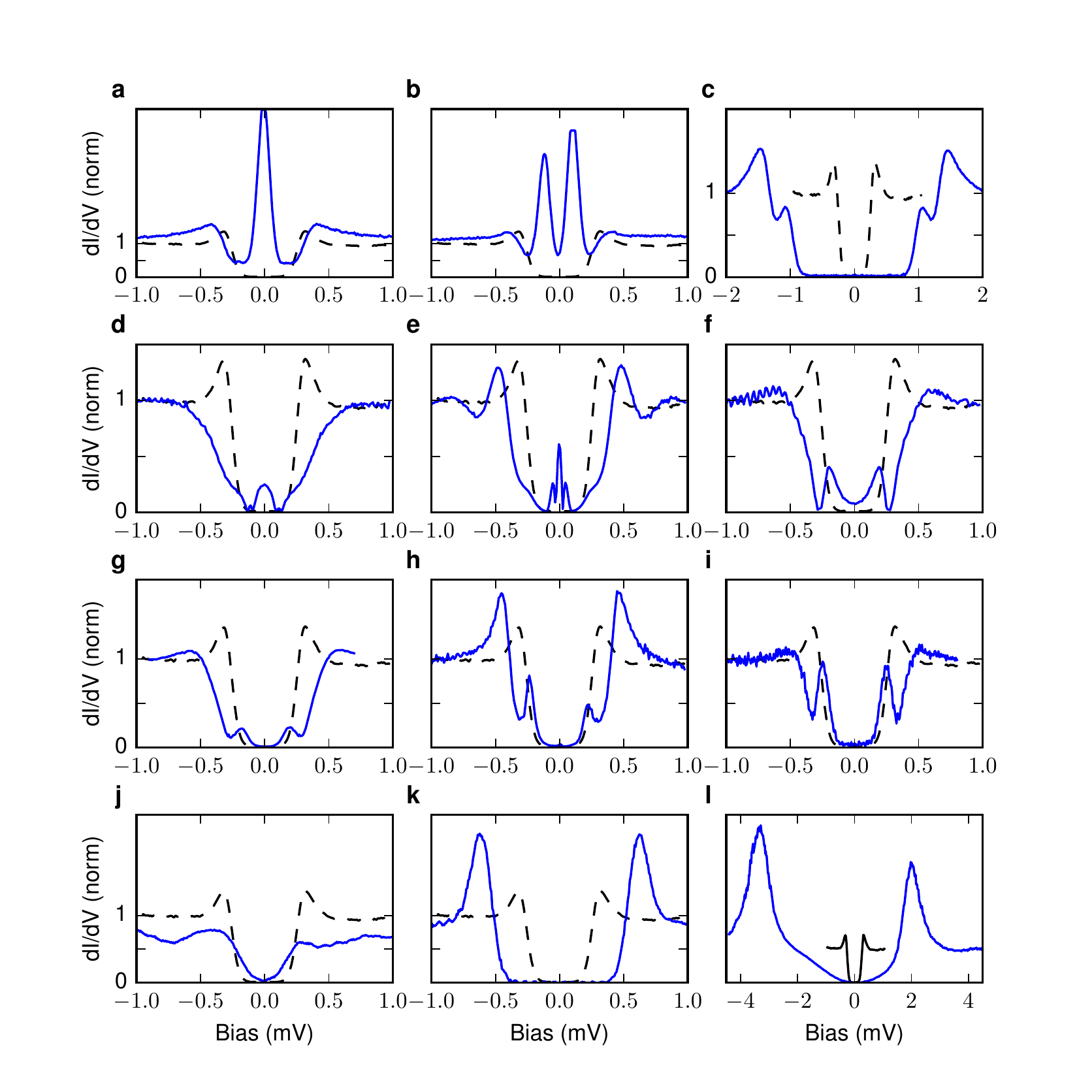}
\caption{\textbf{Examples of spectra observed on different
  Al/EuS/Ag samples (blue) vs. the Al/Ag reference sample (black).} While \textbf{a, b}
can be described with the circuit theory model given in the main article, \textbf{c} features an
enlarged energy gap that cannot be explained with it. \textbf{d-i} have been measured on a second, nominally identical sample. \textbf{d-g} cannot be explained with our current theoretical calculations, but moving to a similar model for very high spin-mixing might allow us to explain these peaks inside the BCS gap in the future
\textbf{h, i} show a triplet gap similar to the one presented in the main article. \textbf{j-l}, spectra found on a
third sample, displaying a variety of different widths of the superconducting energy
gap. Again, as in \textbf{c}, such an increase in the superconducting gap width cannot be
explained by our theoretical model.}
  \label{fig:overview}
\end{figure}

\begin{figure}[htp]
  \centering
  \includegraphics{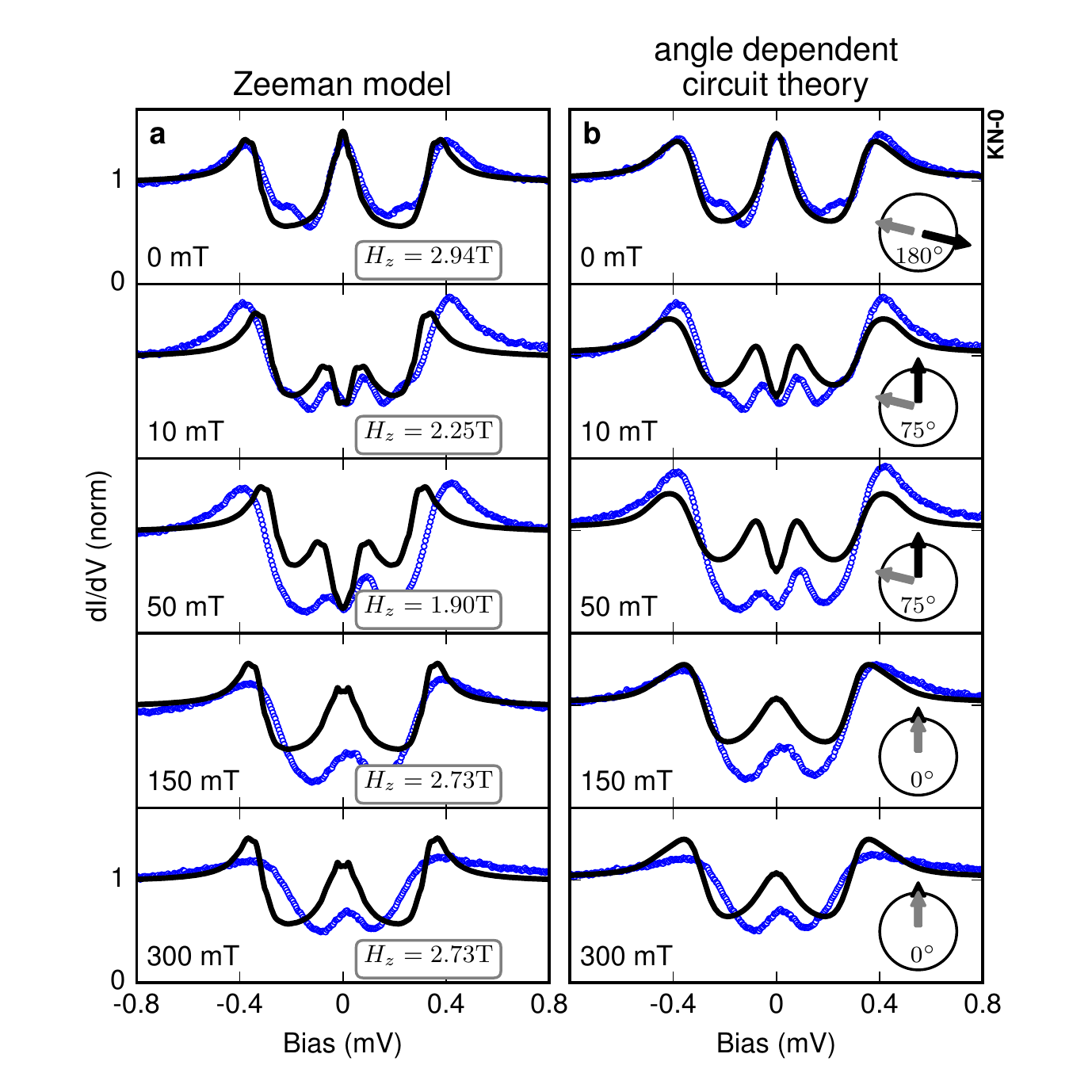}
\caption{\textbf{Comparison of Zeeman-spin-split model and circuit theory model.} Selection of experimental 
$\mathrm{d}I/\mathrm{d}V$ spectra (blue circles) from Fig.~4 of the main text. \textbf{a}, Theoretical modelling with Zeeman-spin-plit LDOS. \textbf{b}, Same fits as shown in Fig.~4 of the main text performed with the circuit theory model. While the spectra can also be described qualitatively by the Zeeman model, the resulting fit parameters show a non-monotonous dependence on the field strength. Data was recorded on sample EuS-1.}
  \label{fig:si-zeeman}
\end{figure}

\begin{table}[htp]
\footnotesize
\begin{tabular}{l c c c c c}
\textbf{\#} 
& EuS-1%KN-0 
& EuS-2%KN-1 
& EuS-3%KN-2 
& EuS-4%KN-3 
& EuS-5%KN-6
\\[5pt]
\hline \\[-5pt]
$p_0$ (mbar) &-& \SI{6e-8}{} & \SI{3e-9}{} &\SI{2e-8}{} &
\SI{5e-9}{} \\
substrate & - & - & \SI{120}{\celsius},& \SI{300}{\celsius}, & \SI{300}{\celsius},\\[-7pt]
treatment & - & - &  \SI{30}{\min} &  \SI{30}{\min}& \SI{30}{\min}\\[7pt]
%--
\textbf{layer 1}  & Al & Al & Al& Al& Al \\
thickness $d_1$ (nm) & 25& 25& 25& 25& 25\\
duration $t_1$ (s) & - & 200& 720& 720& 300\\
rate $r_1$ (\SI{}{\angstrom \per \second})&- & 1.3& 0.35&0.35 \\
$T_{\mathrm S}$ (\SI{}{\kelvin}) & LN2 & LN2 & 110 & 77-85&  110\\
$p_1$ (mbar)&- &\SI{3e-6}{} &\SI{8e-7}{} & - & \SI{6e-8}{}\\[7pt]
%--
\textbf{layer 2}  & EuS & EuS & EuS& EuS& EuS\\
thickness $d_2$ (nm) & 5& 0,2,5,10& 0,5,7,10& 0,5,7,10& 0,2,3.5,5\\
duration $t_2$ (s) & - & 760& 1290& 1050& 840\\
rate $r_2$ (\SI{}{\angstrom \per \second})&- & 0.07& 0.08&0.1 & 0.1\\
$T_{\mathrm S}$ (\SI{}{\kelvin}) & no LN2 & no LN2 & 400 & 400 & 300\\
$p_2$ (mbar)&- &\SI{6e-6}{} &\SI{6e-6}{} & \SI{5e-6}{} & \SI{2e-6}{} \\[7pt]
%--
\textbf{layer 3}  & Ag & Ag & Ag & Ag& Ag\\
thickness $d_3$ (nm) & 5& 5& 5& 5& 5\\
duration $t_3$ (s) & - & 350& 420& -& 330\\
rate $r_3$ (\SI{}{\angstrom \per \second})&- & 0.14& 0.12&- & 0.2-0.3 \\
$T_{\mathrm S}$ (\SI{}{\kelvin}) & no LN2 & no LN2 & 300 & 85 &  300\\
$p_3$ (mbar)&- &\SI{7e-7}{} &\SI{5e-7}{} & \SI{5e-8}{} & \SI{3e-7}{} \\[7pt]
\textbf{TEM} & -& after & -& after & after \\[-7pt]
& & 9 months & & 2.5 months & 12 days\\
\end{tabular}
\caption{\textbf{Summary of the fabrication parameters of the EuS samples.} Note the
  addition of an electric heater to the sample holder between the fabrication of EuS-2 and
  EuS-3, which allows the deposition of the EuS film at \SI{400}{\kelvin}}
    \label{tab:fabrication}
\end{table}

\begin{table}[htp]
\centering
\begin{tabular}{l c c c c}
     &units&~\ref{fig:si-fieldsweep}a &~\ref{fig:si-fieldsweep}b &~\ref{fig:si-fieldsweep}c \\
     & & &~\ref{fig:si-waterfall}a-d &~\ref{fig:si-waterfall}e-h\\
     \hline
     $G_{\mathrm{S}}/(G_{\mathrm{T}}\epsilon_{\mathrm{Th,S}})$ & $(k_{\mathrm{B}}T_{\mathrm{c}})^{-1}$ & 4.1 & 1.25 & 0.5\\
     $G^\phi_{\mathrm{S}}/G_{\mathrm{T}}$&1& 5 & 1.3 & 0.45 \\
     $G_{\mathrm{N}}/(G_{\mathrm{T}}\epsilon_{\mathrm{Th,N}}) $&$(k_{\mathrm{B}}T_{\mathrm{c}})^{-1}$ & 0.07 & 0.01 & 0.035 \\
     $G^\phi/G_{\mathrm{T}}$ &1& -0.061 & -0.05 & - 0.01\\
     $P_n$ &1& 0.6 & 0.6& 0.4\\
     $T_{\mathrm{c}}$ &K& 1.84 & 1.84 & 1.84\\
     $T_{\mathrm{exp}}$ &mK& 310 & 275 & 330
\end{tabular}
\caption{\textbf{Parameters for the theoretical curves found in Supplementary Fig.~\ref{fig:si-fieldsweep} and Supplementary Fig.~\ref{fig:si-waterfall}.}}
\label{tab:parameters}
\end{table}
\clearpage
  
\section{Supplementary Note: Spectroscopic features of S/F/N multilayers}
\label{si:sts}

Two examples for the spectra produced by scanning tunneling spectroscopy (STS) are shown in Supplementary Fig.~\ref{fig:reference}. The measurements were recorded on two different Al films capped with \SI{5}{\nano \meter} Ag, one Al film \SI{244}{\nano \meter} (gray) and the other \SI{25}{\nano \meter} (blue) thick. The recorded $\mathrm{d}I/\mathrm{d}V$ spectra clearly follow the classic shape of the superconducting density of states (DOS), including the BCS gap with its delimiting coherence peaks. The thin film spectrum shown here will serve as a reference sample for all our measurements on Al based superconducting multilayers. Note, that the theory curves shown here are not based on the circuit theory model displayed in later chapters, but on the Usadel equation alone.

Adding a layer of EuS between the Al and Ag results in several significant changes to the $\mathrm{d}I/\mathrm{d}V$ spectra recorded by STS measurements. The first difference between spectra with and without a ferromagnetic layer is the appearance of free states inside the gaps, the ``zero-bias peak'' and ``triplet gap'' features described in the main article. The second difference is an increase of the apparent gap width in some cases by up to several hundred \SI{}{\micro \electronvolt} (see Supplementary Fig.~\ref{fig:5nm-vs-ref}). Comparing with literature \cite{court2008} on single Al layers (not sandwiched with other layers), the apparent gap size in the Al/Ag layer is smaller because of the strong (negative) proximity effect. In the Al/EuS/Ag system the proximity effect is not so strong, because of the ferromagnetic insulator EuS decoupling the superconductor from the normal metal. Also, the effective thickness of the superconducting Al layer, i.e. the fraction of the layer that becomes superconducting, is reduced compared to the geometrical thickness due to the presence of the ferromagnetic EuS. Both effects would give rise to an enhanced gap size in the Al/EuS/Ag system.
%We argue that the presence of the ferromagnetic layer further reduces the effective superconducting layer thickness.
Third, the presence
of the ferromagnet reliably increased the perpendicular critical field of the multilayer
by a factor of at least two. While thin film samples of Al and Ag show a second-order
phase transitions to the normal state around out-of-plane fields of \SI{40}{\milli
  \tesla}, a sample including the EuS layer shows the second-order phase transition only
at around \SI{170}{\milli \tesla} (see Supplementary Fig.~\ref{fig:5nm-vs-ref}). When in-plane magnetic 
fields where applied, the maximum set field of the magnet (\SI{500}{\milli \tesla}) was not sufficient to suppress superconductivity, as expected for Al films of this thickness ($\mu_0 H_{\mathrm{c}} \approx \SI{800}{\milli \tesla}$ is reported for \SI{25}{\nano \meter} films \cite{meservey1994}).  The fourth difference is
that the previously listed phenomena are strongly position dependent for all samples with
the EuS interface, while the reference samples consistently show BCS-like spectra
everywhere on the sample.

\section{Supplementary Note: Sample Characterisation}
\label{si:char}

A series of transmission electron microscopy (TEM) measurements on samples from several different production batches where performed to answer these questions. These microscopy investigations were performed between several months (samples EuS-2 and EuS-4) and only days (sample EuS-5) after sample fabrication and STS analysis (yet always immediately after the lamella was cut), and the samples were stored in air between the measurements.

Scanning transmission electron microscopy (STEM) images (Supplementary Fig.~\ref{fig:tem}a) show continuous film thicknesses for EuS (dark) and Al,
and two oxide layers (bright) above and below the Al film. The contrast within the Al film is a consequence of different crystallites being imaged in different directions with respect to their lattice planes. 
EDX measurements (see Supplementary Fig.~\ref{fig:edx})
show a clean Al film, but superposition of several components especially at the interfaces between the Al and
the EuS layer. Some of these components (Pt and C, the latter being not shown in the EDX analysis) have only been deposited shortly before the focused ion
beam (FIB) cutting of the lamella, but we cannot exclude that the EuS film experienced oxidation and some 
unintended doping during the fabrication process or later, as mentioned above.

High-resolution transmission electron microscopy (HRTEM) images 
(see Supplementary Fig.~\ref{fig:tem}b) show a certain roughness of this oxide layer resulting in a wavy shape of the EuS and the Ag film on top. Also, lattice planes are visible for all layers when fast Fourier transformation (FFTs) analysis is performed on single grains of the film confirming the polycrystalline growth. While the number of visible reflections in the FFT is low, the peaks of highest intensity (for Al and Ag the reflections of the (111) planes, for EuS the reflections from the (200) planes) can be clearly identified in Supplementary Fig.~\ref{fig:tem}c. The oxide layer shows four prominent reflections indicating a lattice constant of \SI{0.296}{\nano \meter}, which fits closely to the high intensity reflections from the (111) plane of cubic EuO or the (402) plane of monoclinic Eu$_2$O$_3$. Because the lattice constants for these compounds are very similar, a closer identification is difficult, but the oxides of Al can be excluded from the list of possible materials.  If it was oxidation of the Al, we would expect a fading of the contrast into the Al layer, what we do not observe. 
Contrarily we observe a sharp contrast change between the Al and the unknown layer and a fading between the unknown layer and the EuS layer. This observation gives an additional indication that the unknown layer is an oxid-containing modification of the EuS layer.
While we can cannot easily distinguish EuO and Eu$_2$O$_3$ in our TEM analysis, the ferromagnetic nature of EuO should clearly separate it from the non-magnetic properties of Eu$_2$O$_3$, which we will show by SQUID magnetometry, revealing a transition around \SI{63}{\kelvin} in addition to the one from the EuS, observed around \SI{17}{\kelvin}. The bulk Curie temperature of EuO is $\approx \SI{69}{\kelvin}$, which is expected to decrease in thin films and small grains.

As explained in the main article, we assume that a magnetically harder (relative to the bulk of EuS) interface magnetisation is responsible for the spectroscopic anomalies we observe in our trilayer films. The possible occurrence of ferromagnetic EuO at the interface could provide a possible source for such a magnetic configuration.

The ferromagnetic layer was characterised by SQUID measurements on a 5-nm-thick EuS film, which show that the ferromagnetic transition appears 
%  around \SI{15}{\kelvin}
(see Supplementary Fig.~\ref{fig:squid-kn0}b), 
%slightly smaller than 
close to the bulk Curie temperature $T_{\mathrm{Curie}} = \SI{16.7}{\kelvin}$. The coercive field of the entire magnetic layer is $\mu_0 H_{\mathrm c} \approx \SI{5}{\milli \tesla}$ and the material is magnetically soft, leading to a shallow magnetisation loop (Fig~\ref{fig:squid-kn0}a, blue). The saturation magnetisation was around 3$\mu_{\mathrm{B}}$ per formula unit, i.e., smaller than the expected value of $7 \mu_{\mathrm B}$ for a completely saturated film. The fact that the magnetisation curve (black) from zero-field cooled (ZFC) state starts close to $M(H = 0) = 0$ hints at a very uniform domain structure with neighbouring magnetic moments that often cancel each other out, i.e., a distribution with no predominant direction of magnetisation. Hence, a sample starts from a demagnetised state and the domains can be rotated easily and quickly with a relatively low magnetic field. 
Supplementary Fig.~\ref{fig:squid-kn1} shows equivalent measurements performed on sample EuS-2 after the STS measurements. The hysteresis loop recorded at \SI{30}{\kelvin} shows an only slightly smaller coercive field than the one at \SI{5}{\kelvin}, indicating the existence of a ferromagnetically ordered phase at that temperature. The temperature dependent measurements show in Supplementary Fig.~\ref{fig:squid-kn1}b indicate a transition occurring $\approx \SI{63}{\kelvin}$, somewhat lower than the Curie temperature of bulk EuO and an increase of the magnetisation below \SI{20}{\kelvin}, consistent with the transition temperature of EuS. These findings support the existence of two magnetically different materials in the sample. From the ratio of the magnetisations we estimate that the material with the higher transition temperature has a much lower volume and might be present in the interface between Al and EuS.

The Al(\SI{25}{\nano \meter})/EuS(\SI{5}{\nano \meter})/Ag(\SI{5}{\nano \meter})
multilayer has a critical temperature of $T_{\mathrm c} \approx \SI{1.7}{\kelvin}$ (see Supplementary Fig.~\ref{fig:temp}),
similar to the critical temperature of the Al(\SI{25}{\nano \meter})/Ag(\SI{5}{\nano
  \meter}) reference sample. When recording a temperature dependence on a $\mathrm{d}I/\mathrm{d}V$ spectrum with a zero-bias peak, the anomaly vanishes at around \SI{700}{\milli \kelvin}.

For more details on fabrication of EuS nanostructures and for more characterisation studies see \cite{wolf2014}.

\section{Supplementary Note: Discussion of the oxide layer}
\label{si:oxide}

As described in the discussion of our TEM results, an oxide barrier was found between the Al and the EuS layer. If this layer was an insulating, nonmagnetic material, it should significantly reduce the electronic coupling between the Al and Ag layers. In order to investigate the effect of such a low conductance of the tunnelling connector, we performed calculations of the LDOS for different values of the tunnelling conductance $G_{\mathrm T}$. Supplementary Fig.~\ref{fig:gt-waterfall} shows the results of this study. As visible from those theory curves, lowering of $G_{\mathrm T}$ first suppresses the features inside the gap, and finally significantly suppresses superconductivity in the Ag layer. This is the expected behaviour for a proximity-coupled system when the coupling strength is decreased. Hence, the existence of a tunnel barrier between the superconductor and the ferromagnetic insulator cannot explain the appearance of pronounced subgap structures, like the once we observe experimentally. From this result and the fact that we do observe subgap features, we conclude that the oxide layer does not form a strong tunneling barrier, but shares the conductance properties of EuS.

\section{Supplementary Note: Magnetic dependence of a triplet gap feature}
\label{si:triplet-gap}
In the main article, the evolution of a zero bias peak was shown and described with the help of our circuit theory model. Figure \ref{fig:si-double-peak-fieldsweep} displays a magnetic dependence of the differential conductance when the tip is positioned above a region like the one shown in Supplementary Fig.~3d. According to our model, the triplet gap hallmarks the local magnetic moments being oriented noncollinearly (a) with respect to each other. As in the main article, an external magnetic field can be used to first reorient the soft magnetic moments of the bulk (b), and at higher fields, the magnetically harder moments at the S/F interface (c). As displayed in (d)-(m), the spectra follow the same trend as in the main article: the triplet gap evolves into a zero-bias peak as the external field increases, which finally vanishes as the external field begins suppressing the superconductivity of the multilayer.

However, there is a significant difference to the measurement presented in the main article: for this magnetic dependence, an out-of-plane magnetic field was used instead of an in-plane field. Because of that, the sample cannot remain magnetised after the external field is switched off, as the magnetic shape anisotropy does not allow the thin film to remain magnetised in an out-of-plane direction. Thus, as the magnetic field is decreased, the magnetic moments in the thin film layer relax back into an in-plane direction of magnetisation.

The magnetic dependence in Supplementary Fig.~\ref{fig:si-double-peak-fieldsweep}d-m shows just that behaviour: the spectra are plotted in chronological order, i.e., after the zero-bias peak in (h) was observed at an external field of \SI{20}{\milli \tesla}, we decreased the field again to \SI{16}{\milli \tesla} and observed the reappearance of the triplet gap in (f). When the field was increased again, suppression of the superconducting state quickly set in at around \SI{30}{\milli \tesla}, as expected for out-of-plane fields. No zero-bias peak was encountered, and decreasing the field again now revealed a standard BCS-like gap. This observation again can be explained by the thin film magnetisation not remaining in the out-of-plane direction. Once the external field is removed, all magnetic moments relax into an in-plane direction randomly. This effect results in the formation of many domains, many of which are smaller than they were in the ZFC state. Since these domains don't share an predominant direction of magnetisation, no significant net magnetisation of the probed region remains, and we observe no subgap structure in the tunnel spectrum.

\section{Supplementary Note: Symmetry of the magnetic dependence measurements}
\label{si:symmetry}

In the main article, a magnetic field dependence measurement with fields in the positive $x$-direction is shown (Fig.~4). In the beginning of the experimental work, we routinely swept the field in both polarities, but did not find significant differences and stopped these systematic control measurements after a while to save time. Supplementary Fig.~\ref{fig:hysteresis} displays an example showing both field directions for another contact. No difference between positive and negative polarity and no hysteresis was observed.

\section{Supplementary Note: Numerical parameters}
\label{si:sim}

Because the parameter space is large, the theoretical curves presented in the main text
are not the only combination of parameters that fit the experimental data well. In
Supplementary Fig.~\ref{fig:si-fieldsweep} we show three sets of curves with different sets of parameters. The calculations based
on the circuit theory were carried out with the parameters shown in Supplementary Tab.~\ref{tab:parameters}. The parameter set resulting in the 
spectra in Supplementary Fig.~\ref{fig:si-fieldsweep}a is the data set resulting in the theoretical curves in Fig.~2 and 
4 of the main article, Supplementary Fig.~\ref{fig:si-fieldsweep}b,c show completely different parameter sets for comparison. 
The lower $G_{\mathrm{S}}$ and $G_{\mathrm{S}}^{\phi}$ terms result in less pronounced zero-bias peaks, but fit the experimental data better once the 
triplet gap appears. In order to fit the experimental data at \SI{300}{\milli \tesla}, where the external magnetic field results in a significant 
decrease of the height of the coherence peaks and in a decrease in gap width, several approaches lead to a better fit:  introducing increased thermal smearing (by increasing the 
temperature of the environment), a lower ratio of $G_{\mathrm{S}}/(G_{\mathrm{T}} \epsilon_{\mathrm{Th,S}})$ (shown in Supplementary Fig.~\ref{fig:si-fieldsweep}c) or a  stronger proximity effect (higher ratio of $G_{\mathrm{N}}/(G_{\mathrm{T}} \epsilon_{\mathrm{Th,N}})$ all describe the experimental data better for that external magnetic field value.

When changed individually, the parameters in Supplementary Tab.~\ref{tab:parameters} have the following effects on the calculated LDOS (all statements are general trends, changing one parameter always influences many features of the spectra and usually necessitate adjusting other parameters): The polarisation $P$ changes the width of the triplet gap and the zero-bias peak, with lower $P$ resulting in a wider gap and a narrower zero-bias peak. $G_{\mathrm{N}}/(G_{\mathrm{T}}\epsilon_{\mathrm{Th,N}})$ can be understood as a measure for the strength of the proximity effect, where a higher value decreases the gap width and scales down the height of the coherence peaks. $G_{\mathrm{S}}/(G_{\mathrm{T}}\epsilon_{\mathrm{Th,S}})$ and the two spin-mixing terms $G^\phi_{\mathrm{S}}/G_{\mathrm{T}}$ and $G^\phi/G_{\mathrm{T}}$ change the nature of the subgap features entirely. When $G_{\mathrm{S}}/(G_{\mathrm{T}} \epsilon_{\mathrm{Th,S}})$ alone is increased, the width of the triplet gap increases and the amplitude of the zero-bias peak decreases. When $G_{\mathrm{S}}^{\phi}/G_{\mathrm{T}}$ is increased, the zero-bias peak increases in amplitude. When $G_{\mathrm{S}}$ and $G^{\phi}_{\mathrm{S}}$ are increased simultaneously, both the features inside the gap and the coherence peaks increase in amplitude. When $G^{\phi}$ is decreased, the spectra for $\theta = 0$ and $\theta = \pi$ become more similar and the $\theta$ dependence gets increasingly more symmetric.

For all calculations, the superconducting node has a critical temperature $T_{\mathrm{c}} = \SI{1.84}{\kelvin}$, which self-consistently results in a superconducting energy 
gap $\Delta= 1.764 k_{\mathrm B} T_{\mathrm{c}} = \SI{280}{\micro \electronvolt}$. The experimental temperature $T_{\mathrm{exp}}$ was set to between \SI{275}{\milli \kelvin} 
and \SI{330}{\milli \kelvin} in order to achieve the amount of thermal smearing required for a good fit to the experimental data. 
Both $T_{\mathrm{c}}$ and $T_{\mathrm{exp}}$ are realistic for thin film Al ($T_{\mathrm{c}}$ measurement in Supplementary Fig.~\ref{fig:temp}) and the cryostat (an Oxford Instruments Heliox$^{\textsc{TM}}$ VL) used for our experiments.

The pair amplitudes of the resulting triplet states are shown in Supplementary Fig.~\ref{fig:si-waterfall} for the parameter combinations shown in Supplementary Fig.~\ref{fig:si-fieldsweep}b,c. The evolution of the
triplet pairing amplitudes behaves in the same way for those two parameter sets as they do for the set used in the main article (Fig.~2), showing that although we cannot make 
profound predictions about the spin-dependent parameters present in the S/FI interface in our experiments, the message stays the same: there is a direct
correlation between the opening of a subgap inside the superconducting energy gap and the presence of equal-spin triplet Cooper pairs.

\section{Supplementary Note: other features inside the gap}

Supplementary Fig.~\ref{fig:overview} shows a variety of subgap features measured on different
Al(\SI{25}{\nano \meter})/EuS(\SI{5}{\nano \meter})/Ag(\SI{5}{\nano \meter}) samples. The
spectra depicted in Supplementary Fig.~\ref{fig:overview}a and b are of the kind described in the main
article. They fit our theory model well. Supplementary Fig.~\ref{fig:overview}d-i were recorded on a different Al(\SI{25}{\nano
    \meter})/EuS(\SI{5}{\nano\meter})/Ag(\SI{5}{\nano \meter}) sample, and while those spectra also show zero-bias peaks (d,e) and 
    an additional gap-like feature (f,i), these spectra have an important difference to the spectra shown in the main text: the differential conductance 
    reaches $\mathrm{d}I/\mathrm{d}V$ = 0 or at least $\mathrm{d}I/\mathrm{d}V \ll 0.5 G_{\mathrm{b}}$ inside the gap at energies higher than the edge of the feature. Although we probed a substantial part of the parameter space, for none of the tested parameter combinations, this behaviour could be obtained from our theoretical model. We currently have no 
    explanation for these features, and will hopefully be able to model them in future work via a similar circuit theory model for very high spin-mixing terms.
    
    For the the type of spectra shown in Supplementary Fig.~\ref{fig:overview}c,h,i, theory curves based on a simple Zeeman field spin-splitting the BCS gap \cite{meservey1994} can also result in a good quality of fit. Such a model would not explicitly require the creation of equal-spin triplet pairs. In the case of the spectra displayed in Fig~\ref{fig:overview}h,i, the Zeeman-splitting theory also fails to explain the low $\mathrm{d}I/\mathrm{d}V$ values at energies outside the subgap feature. Summarising, our circuit theory model fits well to all spectra, i.e. those that can and those that cannot be  described by a Zeeman splitting of the density of states.
    
    The spectra in Supplementary Fig.~\ref{fig:overview}c,k,l show a large increase in gap width, we observe $\Delta \approx \SI{1.2}{\milli \electronvolt}$, 
    $\Delta \approx \SI{0.5}{\milli \electronvolt}$ and $\Delta \approx \SI{5.5}{\milli \electronvolt}$, respectively.
    These gap energies are surprising, and we will attempt to explain them in a forthcoming publication.

\section{Supplementary Note: Attempt to fit the spectra of Fig.~4 with a Zeeman-model}
\label{si:zeeman}

Supplementary Fig.~\ref{fig:si-zeeman} shows the results of fitting the spectra displayed in Fig.~4 on the main text with a model that assumes a Zeeman-splitted LDOS along the lines of Ref. \cite{meservey1994}. To this end, we first fit dI/dV of the Ag/Al sample (see Supplementary Fig.~\ref{fig:reference}). To mimic a spin-split LDOS, we then superpose two of these fitted spectra shifted with respect to each other. Fitting parameters are the relative height of the two parts as well as their splitting. We then translate the calculated splitting into a field strength assuming a g-factor of 2. While qualitatively the  shape of the individual spectra can be reproduced with this model, the fitting parameters depend non-systematically on the applied field.